\documentclass[12pt,nohyper]{JHEP3}
\usepackage{graphics,psfrag}
\usepackage{amsthm,amssymb,epsfig,amsmath,euscript,array,cite}

\newcounter{multieqs}

\newcommand{\be}{\begin{equation}}
\newcommand{\ee}{\end{equation}}
\newcommand{\eq}[1]{(\ref{#1})}

\newcommand{\bm}[1]{\mbox{\boldmath $#1$}}
\newcommand{\rf}[1]{(\ref{#1})}

\def\bd{\begin{document}}
\def\ed{\end{document}}
\def\nn{\nonumber}
\def\bea{\begin{eqnarray}}
\def\eea{\end{eqnarray}}
\let\bm=\bibitem
\let\la=\label

\def\npb#1#2#3{Nucl. Phys. {\bf{B#1}} #3 (#2)}
\def\plb#1#2#3{Phys. Lett. {\bf{#1B}} #3 (#2)}
\def\prl#1#2#3{Phys. Rev. Lett. {\bf{#1}} #3 (#2)}
\def\prd#1#2#3{Phys. Rev. {D \bf{#1}} #3 (#2)}
\def\cmp#1#2#3{Comm. Math. Phys. {\bf{#1}} #3 (#2)}
\def\cqg#1#2#3{Class. Quantum Grav. {\bf{#1}} #3 (#2)}
\def\nppsa#1#2#3{Nucl. Phys. B (Proc. Suppl.) {\bf{#1A}}#3 (#2)}
\def\ap#1#2#3{Ann. of Phys. {\bf{#1}} #3 (#2)}
\def\ijmp#1#2#3{Int. J. Mod. Phys. {\bf{A#1}} #3 (#2)}
\def\rmp#1#2#3{Rev. Mod. Phys. {\bf{#1}} #3 (#2)}
\def\mpla#1#2#3{Mod. Phys. Lett. {\bf A#1} #3 (#2)}
\def\jhep#1#2#3{J. High Energy Phys. {\bf #1} #3 (#2)}
\def\atmp#1#2#3{Adv. Theor. Math. Phys. {\bf #1} #3 (#2)}

\def\N{{\cal N}}
\def\sst{\scriptscriptstyle}
\def\thetabar{\bar\theta}
\def\Tr{{\rm Tr}}
\def\one{\mbox{1 \kern-.59em {\rm l}}}

\def\a{\alpha}      \def\da{{\dot\alpha}}  \def\dA{{\dot A}}
\def\b{\beta}       \def\db{{\dot\beta}}  
\def\g{\gamma}  \def\G{\Gamma}  \def\dc{{\dot\gamma}}  
\def\d{\delta}  \def\D{\Delta}  \def\ddt{\dot\delta}  
\def\e{\epsilon}        \def\ve{\varepsilon}  
\def\f{\phi}    \def\F{\Phi}    \def\vvf{\f}  
\def\h{\eta}  
\def\k{\kappa}  
\def\l{\lambda} \def\L{\Lambda}  
\def\m{\mu} \def\n{\nu}  
\def\o{\omega}  
\def\p{\pi} \def\P{\Pi}  
\def\r{\rho}  
\def\s{\sigma}  \def\S{\Sigma}  
\def\t{\tau}  
\def\th{\theta} \def\Th{\Theta} \def\vth{\vartheta}  
\def\X{\Xeta}  
\def\z{\zeta}  

\def\na{\nabla}  

\def\cA{{\cal A}} \def\cB{{\cal B}} \def\cC{{\cal C}}  
\def\cD{{\cal D}} \def\cE{{\cal E}} \def\cF{{\cal F}}  
\def\cG{{\cal G}} \def\cH{{\cal H}} \def\cI{{\cal I}}  
\def\cJ{{\cal J}} \def\cK{{\cal K}} \def\cL{{\cal L}}  
\def\cM{{\cal M}} \def\cN{{\cal N}} \def\cO{{\cal O}}  
\def\cP{{\cal P}} \def\cQ{{\cal Q}} \def\cR{{\cal R}}  
\def\cS{{\cal S}} \def\cT{{\cal T}} \def\cU{{\cal U}}  
\def\cV{{\cal V}} \def\cW{{\cal W}} \def\cX{{\cal X}}  
\def\cY{{\cal Y}} \def\cZ{{\cal Z}}

\def\ua{\underline{\alpha}}  
\def\uc{\underline{\phantom{\alpha}}\!\!\!\gamma}  
\def\um{\underline{\mu}}  
\def\ud{\underline\delta}  
\def\ue{\underline\epsilon}  
\def\una{\underline a}\def\unA{\underline A}  
\def\unb{\underline b}\def\unB{\underline B}  
\def\unc{\underline c}\def\unC{\underline C}  
\def\und{\underline d}\def\unD{\underline D}  
\def\une{\underline e}\def\unE{\underline E}  
\def\unf{\underline{\phantom{e}}\!\!\!\! f}\def\unF{\underline F}  
\def\unm{\underline m}\def\unM{\underline M}  
\def\unn{\underline n}\def\unN{\underline N}  
\def\unp{\underline{\phantom{a}}\!\!\! p}\def\unP{\underline P}  
\def\unq{\underline{\phantom{a}}\!\!\! q}  
\def\unQ{\underline{\phantom{A}}\!\!\!\! Q}  
\def\unH{\underline{H}}  
  
\def\As {{A \hspace{-6.4pt} \slash}\;}  
\def\bs {{b \hspace{-6.4pt} \slash}\;}  
\def\Ds {{D \hspace{-6.4pt} \slash}\;}  
\def\ds {{\del \hspace{-6.4pt} \slash}\;}  
\def\ss {{\s \hspace{-6.4pt} \slash}\;}  
\def\ks {{ k \hspace{-6.4pt} \slash}\;}  
\def\ps {{p \hspace{-6.4pt} \slash}\;}   
\def\xs {{x \hspace{-6.4pt} \slash}\;}  
\def\pas {{{p_1} \hspace{-6.4pt} \slash}\;}  
\def\pbs {{{p_2} \hspace{-6.4pt} \slash}\;}   
\def\cFs {{{\cal F} \hspace{-6.4pt} \slash}\;}

\def\Dh{\hat{D}}
\def\Gh{\hat{G}}
\def\Fh{\hat{F}}
\def\Ph{\hat{P}}
\def\Rh{\hat{R}}
\def\Vh{\hat{V}}  
\def\Xh{\hat{X}} 
 
\def\ah{{\hat{a}}}
\def\gh{\hat{g}} 
\def\hh{\hat{h}}
\def\uh{\hat{u}}  
\def\xh{\hat{x}}  
\def\yh{\hat{y}}  
\def\ph{\hat{p}}  
\def\xih{\hat{\xi}}  
\def\chih{\hat{\chi}}

\def\psit{\tilde{\psi}}  
\def\Psit{\tilde{\Psi}}   
\def\Psibt{\tilde{\bar{Psi}}}  

\def\st{\tilde{\sigma}}  

\def\Phit{\tilde{\Phi}}   
\def\Phitb{\overline{\tilde{Phi}}}  
\def\tht{\tilde{\th}}  
\def\lt{\tilde{\l}}
\def\chit{\tilde{\chi}}   
\def\phit{\tilde{\phi}} 

\def\At{\tilde{A}}
\def\Bt{\tilde{B}}
\def\Ct{\tilde{C}}
\def\Dt{\tilde{D}}
\def\Et{\tilde{E}}
\def\Ft{\tilde{F}}
\def\Ht{\tilde{H}}
\def\Qt{\tilde{Q}}  
\def\Rt{\tilde{R}}  
\def\Mt{\tilde{M }}  
\def\Nt{\tilde{N}}   
\def\St{\tilde{S}}
\def\Vt{\tilde{V}}
\def\Xt{\tilde{X}} 
\def\at{\tilde{a}}
\def\ct{\tilde{c}}   
\def\htt{\tilde{h}} 
\def\ft{\tilde{f}}
\def\gt{\tilde{g}}
\def\pt{\tilde{p}}  
\def\qt{\tilde{q}}  
\def\vt{\tilde{v}}  
\def\nt{\tilde{n}}  
\def\ut{\tilde{u}}  
\def\wt{\tilde{w}}  
\def\zt{\tilde{z}} 
\def\xt{\tilde{x}} 
\def\yt{\tilde{y}} 
\def\Psit{\tilde{\Psi}}
\def\vphit{\tilde{\varphi}}  

\def\delb{\bar{\partial}}  
\def\thb{\bar{\theta}}
\def\mub{\bar{\mu}}
\def\lamb{\bar{\l}}
\def\psib{\bar{\psi}}
\def\sb{\bar{\sigma}}
\def\xib{\bar{\xi}}
\def\chib{\bar{\chi}}

\def\Phib{\bar{\Phi}}
\def\Lamb{\bar{\Lambda}}
\def\Sb{{\overline \Sigma}}
\def\cb{\bar{c}}
\def\wb{\bar{w}}
\def\ub{\bar{u}}
\def\zb{{\bar{z}}}
\def\Qb{{\bar Q}}
\def\qb{\bar{q}}

\def\Ab{{\overline A}} \def\Bb{{\overline B}} \def\Cb{{\overline C}}  
\def\Db{{\overline D}} \def\Eb{{\overline E}} \def\Fb{{\overline F}}  
\def\Gb{{\overline G}} \def\Hb{{\overline H}} \def\Ib{{\overline I}}  
\def\Jb{{\overline J}} \def\Kb{{\overline K}} \def\Lb{{\overline L}}  
\def\Mb{{\overline M}} \def\Nb{{\overline N}} \def\Ob{{\overline O}}  
\def\Pb{{\overline P}}  \def\Rb{{\overline R}}  
 \def\Tb{{\overline T}} \def\Ub{{\overline U}}  
\def\Vb{{\overline V}} \def\Wb{{\overline W}} \def\Xb{{\overline X}}  
\def\Yb{{\overline Y}} \def\Zb{{\overline Z}}  

\def\fb{{\overline f}}
\def\gb{{\overline g}}
\def\mb{{\overline m}}
\def\lb{{\overline l}}
\def\yb{{\overline y}}

\def\ba{{\bf a}} 
\def\bk{{\bf k}}  
\def\bl{{\bf l}}  
\def\bp{{\bf p}}  
\def\bq{{\bf q}}  
\def\br{{\bf r}}
\def\bt{{\bf t}}
\def\bu{{\bf u}}
\def\bv{{\bf v}}
\def\bx{{\bf x}}  
\def\by{{\bf y}}  
\def\bR{{\bf R}}  
\def\bV{{\bf V}}

\def\bone{{\bf 1}}  

\def\va{{\vec a}}
\def\vk{{\vec k}}
\def\vp{{\vec p}}
\def\vq{{\vec q}}
\def\vx{{\vec x}}
\def\vy{{\vec y}}
\def\vu{{\vec u}}
\def\vv{{\vec v}}

\def\vs{{\vec \sigma}}
\def\vtau{{\vec \tau}}

\newcommand{\ov}[1]{\overrightarrow{#1}}
  
\def\d{\delta}\def\D{\Delta}\def\ddt{\dot\delta}  
  
\def\pa{\partial} \def\del{\partial}  
\def\xx{\times}  
\def\uno{\mbox{1 \kern-.59em {\rm l}}}    
  
\def\trp{^{\top}}  
\def\inv{^{-1}}  
\def\dag{{^{\dagger}}}  
\def\pr{^{\prime}}  
  
\def\rar{\rightarrow}  
\def\lar{\leftarrow}  
\def\lrar{\leftrightarrow}  
  
\newcommand{\0}{\,\!}      
\def\one{1\!\!1\,\,}  
\def\im{\imath}  
\def\jm{\jmath}  
  
\newcommand{\tr}{\mbox{tr}}  
\newcommand{\slsh}[1]{/ \!\!\!\! #1}  
  
\def\vac{|0\rangle}  
\def\lvac{\langle 0|}  
  
\def\hlf{\frac{1}{2}}  
\def\ove#1{\frac{1}{#1}}  

\def\Box{\square}  
\def\CC {\mathbb{C}}
\def\RR{\mathbb{R}}
\def\ZZ{\mathbb{Z}}  
\def\bb#1{{\bf #1}}  
\def\bcomment#1{}  
\def\bfhat#1{{\bf \hat{#1}}}  
\def\VEV#1{\left\langle #1\right\rangle}  

\newcommand{\ex}[1]{{\rm e}^{#1}} \def\ii{{\rm i}}  

\newcommand{\lrbrk}[1]{\left(#1\right)}
\newcommand{\sfrac}[2]{{\textstyle\frac{#1}{#2}}}
 
\def\stw{{\sqrt{2}}}

\def\rf {{\rm f}}
\def\ri {{\rm i}}
\def\rs {{\scriptscriptstyle \rm S}}
\def\rt {{\scriptscriptstyle \rm T}}

\def\rQ {{\scriptscriptstyle \rm \cQ}}
\def\rR {{\scriptscriptstyle \rm \cR}}

\def\cQb{{\cal \Qb}}
\def\cRb{{\cal \Rb}}
\def\cWb{{\cal \Wb}}

\def\fd {{\rm N}}
\def\afd {{\overline{\rm N}}}

\def \II {I\hspace{-.1em}I\hspace{.1em}}
\def \IIA {\mbox{\II A\hspace{.2em}}}
\def \IIB {\mbox{\II B\hspace{.2em}}}
\def \gs {g^s}
\def \ls {\lambda^s}

\def \I {{\cal I}}
\def \qs {q\hspace{-.53em}/\hspace{.15em}}
\def \ks {k\hspace{-.53em}/\hspace{.15em}}
\def \YM {{\mbox{\tiny YM}}}
\def \gym {g_{\YM}}

\def \Lc {\L_c}
\def\IR{\relax{\rm I\kern-.18em R}}
\def \id {{\bf 1}}

\author{Chong-Sun Chu and Douglas J Smith \\  
Centre for Particle Theory
and Department of Mathematics, 
Durham University, Durham, DH1 3LE, UK \\
E-mail:  
\email{chong-sun.chu@durham.ac.uk},   
\email{douglas.smith@durham.ac.uk} }

\title {Towards the Quantum Geometry of 
the M5-brane
in a Constant $C$-Field from Multiple Membranes}

\abstract{
We show that the Nahm equation which describes a fuzzy D3-brane in the
presence of a $B$-field can be derived as a boundary condition of the
F1-strings ending on the D3-brane, and that the modifications of the
original Nahm equation by a $B$-field can be understood in terms of the
noncommutative geometry of the D3-brane. Naturally this is consistent
with the alternative derivation by quantising the open strings in the
$B$-field background. 
We then consider a configuration of multiple M2-branes ending on an
M5-brane with a constant 3-form $C$-field.
By analogy with the case of
strings ending on a D3-brane with a constant $B$-field, one can expect that this
system can be described in terms of the boundary of the M2-branes moving
within a certain kind of quantum geometry on the M5-brane worldvolume.
By repeating our analysis, we
show that the analogue of the $B$-field modified 
Nahm equation, 
the $C$-field modified Basu-Harvey equation
can 
also be understood as a
boundary condition of the M2-branes. We then compare this to the
M5-brane BIon description and show that the two descriptions match
provided we postulate a new type of quantum geometry on the M5-brane
worldvolume. Unlike the D-brane case, this is naturally expressed in
terms of a relation between a 3-bracket of the M5-brane worldvolume
coordinates and the $C$-field. }
 
\preprint{DCPT-09/01}
\keywords{Non-Commutative Geometry,  D-Branes, Membranes}

\begin{document}


\section{Introduction and Summary}


One of the most interesting recent developments in string theory
is the discovery of a
description of the conformal field theory describing multiple membranes
\cite{BLG1,BLG2,BLG3,BLG4}. The Bagger-Lambert (BL) theory 
\cite{BLG1,BLG2,BLG3} was originally motivated by
trying to construct an action with manifest ${\cal N} = 8$
superconformal symmetry, based on a BPS equation postulated by Basu and
Harvey \cite{BH}. This naturally led to an action with a non-Abelian
symmetry based on a 3-algebra. Unfortunately this is problematic as
there is only one example of such a 3-algebra, describing 2 M2-branes 
\cite{LT}.
Attempts have been made to circumvent this difficulty by relaxing some
of the constraints on the 3-algebra. E.g.\ relaxing the requirement for
the 3-algebra to have a Euclidean metric turns out to give a consistent
action describing $N$ branes \cite{lor}, but it appears that this gives an
alternative description of the low energy limit of $N$ D2-branes rather
than M2-branes \cite{d2d2}. However, it has been argued that the M2-brane
theory can be described by a non-Abelian twisted Chern-Simons theory
\cite{ABJM} which has manifest ${\cal N} = 6$ superconformal symmetry.
This theory was not described using a 3-algebra, but it has been shown
that there is an equivalent description using a type of 3-algebra
which is not completely anti-symmetric
\cite{n6}. It therefore seems that the 3-algebra is not needed to
describe multiple M2-branes, but it can be used and indeed can be
viewed as a natural way to encode various constraints on the couplings
and matter content of the theory.

The M5-brane is another mysterious object in M-theory. It is somewhat
analogous to a D-brane in string theory, in the sense that M2-branes can
end on it. However, it contains a self-dual 3-form field strength,
rather than the 2-form field strength on D-branes. The self-duality of
this field leads to technical complications, but the action and equations
of motion for a single M5-brane are known \cite{m51,m52,m53}. 
On the other hand, 
the action or equations of motion for multiple M5-branes are not
understood as it is not at all clear how to formulate a theory with a
non-Abelian 3-form field strength. It would certainly be interesting to
understand more about the theory of multiple M5-branes. 

Since M2-branes can end on M5-branes, one may wonder what can be learned
of the M5-brane from the M2-brane by considering such an intersection.
It is known that the 
equations
of motion for a single M5-brane can be
derived by demanding the $\k$-symmetry of the open membrane ending on it
\cite{km}. It is also known that this intersecting configuration can be
described in terms of the M5-brane theory as a BIon spike, with the
M2-branes emerging from the M5-brane worldvolume. Alternatively, it can
be described in terms of multiple M2-branes as a fuzzy funnel, with the
extra 3 worldvolume dimensions of the M5-brane arising as a fuzzy
3-sphere. In fact this latter description is given by a solution of the
Basu-Harvey equation which was consequently proposed as a BPS equation
for the BL theory of multiple M2-branes.

In an analogous but simpler system with a 
string ending on a D-brane, one of the interesting results
that can be obtained from the open string ending on the D-brane is that
the D-brane worldvolume becomes noncommutative when there is a constant NS $B$-field
present on the D-brane worldvolume \cite{CDS,DH,SW}. The result can 
be derived by  quantising the open string ending on the D-brane. Due to
the $B$-field, the usual Neumann boundary condition becomes a mixed one.
A proper quantization taking account of the boundary condition
\cite{CH1,CH2,chu} gives the result that the endpoints of the open strings,
in other words the D3-brane worldvolume coordinates $X^i$ 
become noncommutative and obey
the commutation relation:
\be \label{ncg}
[X^i,X^j] = i \th^{ij},
\ee
where $\th^{ij}$ is an antisymmetric constant matrix whose components
are related to the components of the NS $B$-field \cite{CH1}. 

Naturally one would like to repeat the same steps for an M5-brane with
a constant $C$-field on it and derive the form of quantum
geometry on the M5-brane. In this case the boundary condition is also a
mixed one, but is nonlinear in the boundary coordinates.
The analysis is much more complicated due to the nonlinear nature of 
the membrane action and one can only do an approximate analysis. 
These results were 
expressed in terms of commutators of the boundary string 
coordinates $X^i(\t,\s_1,\s_2=0)$ and look very complicated
\cite{M2-C}. Also it is not
clear how the full exact results may be obtained in a consistent manner
from this approach. 

In this paper we argue   that the quantum geometry over
the M5-brane naturally takes the form
\be \label{ncg-M5}
[X^i,X^j,X^k ] = i \Theta^{ijk},
\ee
where $\Theta^{ijk}$ is a constant completely antisymmetric  matrix
whose components are related to the components of the constant 
$C$-field. This suggests  that the
natural language to encode the quantum  geometry for the
M5-brane in the presence of a constant $C$-field is in terms of a 
3-bracket rather than a commutator.

To start with, we recall the familiar analogous situation in string
theory and re-derive the noncommutative geometry \eq{ncg} using a new
method. 
Our starting point is the observation that the
D3-D1 intersecting system can be looked upon as a single brane where one
kind of brane can be constructed as a solitonic configuration of the
other system of branes: from the D3-brane point of view, the D1-branes
can be described as a magnetic monopole of charge $N$ in the Abelian
Born-Infeld theory of the D3-brane \cite{callan-mal}. In terms of the
D1-strings theory, one can describe the D3-brane as a certain solution
of the Nahm equation 
whose transverse scalar fields are described by a fuzzy two-sphere
\cite{myers1,myers2}. The fuzzy dimensions provide the two extra dimensions to
build up the D3-brane from the D1-strings. 
Note that in this dual description of the D1-D3 intersecting system,  
it is necessary to 
consider a large number $N$ of D1-branes ending on a D3-brane in order
for the description to be valid. 

The construction  of the D3-brane {\it out of} D1-branes is interesting. 
However since the original system we are describing is really a system of 
D1-strings ending on a D3-brane, it suggests that one should be
able to understand the defining Nahm equation as a
boundary condition of the open D1-strings. We show that this 
is indeed the case. The theory we are using here is a matrix theory of
the D1-strings.
This way to derive the Nahm equation thus provides a new way
to understand the intersecting D1-D3 branes system.  This is one of the
main results of this paper.

One can also include a $B$-field in the transverse directions of the D1-strings, 
corresponding to having a $B$-field on the D3-brane. This system has been
studied both as a BIon spike \cite{moriyama} or as a fuzzy funnel \cite{KC}. 
Basically the bunch of
D1-branes was found to have a deformed shape and to be tilted away from
the normal to the D3-brane. These features can be explained by a constant 
shift in the Nahm equation. This modified Nahm equation 
was derived as a BPS equation of the non-Abelian Born-Infeld theory of
the D1-branes \cite{KC}. Using our new understanding of the Nahm
equation, we re-derive this equation as a boundary condition of the
theory of matrix F1-strings ending on the D3-brane. 
In this analysis, this shift can be understood precisely in terms of the
noncommutative geometry \eq{ncg} of the D3-brane.
Conversely if one did not know about the
noncommutative geometry of the D3-brane, one could have derived it this
way by using the fact that the constant shift  
in the Nahm equation is known to be present by matching with the BIon
description with $B$-field. The noncommutative geometry \eq{ncg} was 
originally  
derived by quantising the open string in a background $B$-field. 
Our derivation of the Nahm equation in the presence of a $B$-field thus 
provides a new way to derive \eq{ncg}. This is another main result of
the paper.

Our main interest is the M2-M5 intersecting system. 
Applying the same idea, we show that the
Basu-Harvey equation can be derived as a boundary condition of the
theory of 
multiple M2-branes. We then proceed to include a $C$-field on the
M5 brane. The system has been studied from the M5-brane point of view
where the M2-branes bundle has been constructed as a certain static charge
configuration protruding out of the M5-brane. Due to the $C$-field, the
M2-branes are static only if tilted away from the normal to the
M5-brane. The $C$-field also modifies the shape of the M2-branes funnel. 
It turns out one can
reproduce precisely the tilting and the shape of the M2-branes funnel 
from the M5-brane point of view \cite{michishita,youm} 
if the Basu-Harvey equation is modified 
in a particular way.
We identify and propose this modified Basu-Harvey equation as a
description of M5-branes with $C$-field.

Now just as one can understand the Nahm equation with $B$-field as a
boundary condition of the theory of open F1-strings, and since we have
already shown that the original Basu-Harvey equation can be understood
as a boundary condition of open membranes, it strongly suggests that one
should be able to understand the modified Basu-Harvey equation as a
boundary condition of open membranes in the presence of $C$-field. We
show that this is indeed the case provided that the M5-brane worldvolume
is described by a quantum geometry of the form \eq{ncg-M5}. This is another
main result of the paper. 

The plan of the paper is as follows. In section 2, we first review the
dual descriptions of the D1-D3 intersecting system in terms of the BPS
equations of the corresponding Born-Infeld theories: a
monopole equation for the Abelian Born-Infeld theory for the D3-brane
and a Nahm equation for the non-Abelian Born-Infeld theory for the
D1-strings.  
In section 3, we show that the Nahm equation can be understood as a
boundary condition of the matrix theory of D1-strings which end on the
D3-brane. 
In section 4, we include a NS $B$-field. By using  a system of 
F1-strings to probe the D1-D3 system, we show that
the Nahm equation with $B$-field can be derived as a boundary condition of
the matrix theory of F1-strings. 
In particular we show that
the Nahm equation with $B$-field encodes 
the noncommutative geometry of the D3-brane as well as information about the
open string metric \cite{SW} on the D3-brane.  
In section 5, we apply our idea to the M2-M5 intersecting
system and derive the  Basu-Harvey equation as a boundary condition of
the theory of multiple open membranes. 
We then generalize the Basu-Harvey equation by including a constant
$C$-field by  checking that our modification results in a 
fuzzy-funnel configuration of
M2-branes with an M5-brane at an angle, reproducing the known result from the
M5-brane BIon solution in a constant $C$-field.
We also show that this equation can be understood as the boundary
condition of open membranes ending on the M5-brane with $C$-field
provided that the 
quantum geometry \eq{ncg-M5} of the M5-brane holds. 
The paper 
concludes 
with some further discussions.

\section{Review of Dual Descriptions of  D1-D3-branes Intersections}

Consider a system of $N$ D1-strings ending 
on a D3-brane.
The intersecting brane system can be described in two different ways in
terms of either the D3-brane theory or the D1-strings theory. 

\subsection{The BIon solution of the D3-brane}

From the
D3-brane theory point of view, the 
Abelian Born-Infeld action is
\be \label{BI-d3}
S_{D3}=-T_3 \int d^{4} x \, \sqrt{-\det\left(
\eta_{\m\n} + \del_\m X^A \del_\n X^A + \l (F_{\m\n}+B_{\m\n}) 
\right)
},
\ee
where  $\lambda=2\pi \a'$ and 
\be
T_p = \frac{1}{g_s(2\pi)^p l_s^{p+1}}
\ee
is the tension for D$p$-brane.
Here $\m, \n=0,1,2,3$  are the worldvolume indices of the 
D3-brane, 
$A,B = 4,\cdots, 9$
are the indices of the transverse space, and
$B_{\m\n}$ is the pull-back of the $B$-field to the worldvolume.
One finds that the Born-Infeld theory 
supports solitonic configurations which describe D1-strings protruding 
from the D3-brane \cite{callan-mal}. For a configuration with a single
transverse excitation $X^9(x^i)$, the Born-Infeld theory \eq{BI-d3} 
admits a static BPS equation  of the form:
\be \label{d3-bps}
\del_i X^9 = B_i , \quad i,j,k =1,2,3, 
\ee
where  $B_i = \frac{1}{2} \e_{ijk} F_{jk}$ is the magnetic field on the
D3-brane. This equation coincides with the BPS equation for magnetic monopoles. 
A simple solution is given by 
\be
X^9(x^i) = \frac{Q}{\sqrt{(x^1)^2+(x^2)^2 +(x^3)^2}}, \qquad 
B_i = -\frac{N}{2r^3}\vec{r}
\ee
where $Q := \pi \a' N$, $N$ is an integer. This corresponds to placing 
$N$ units of $U(1)$ magnetic charge on the D3-brane
and  
describes a spike of $N$ D1-strings coming out of the D3-brane.  
It is interesting to note that at a fixed distance $X^9 =\s$ from the
D3-brane, the cross-section of the D3-brane is a 2-sphere with  radius
\be \label{bion-0}
r(\s) = \frac{ Q}{\s}.
\ee

One can also include a constant NS $B$-field on the worldvolume of the
D3-brane. Let's take a $B$-field in the 12-direction, $B_{12} \neq 0$. 
Due to the $B$-field, the spherical symmetrical is broken. It was found
that \cite{moriyama} the solution is modified to one whose
cross section becomes an ellipsoid  
\be \label{bion-1}
\frac{x_1^2}{r_1(\s)^2}+ \frac{x_2^2}{r_2(\s)^2}+
\frac{(x_3- \s \tan \a)^2}{r_3(\s)^2}
=1,
\ee
where 
\be \label{radii}
r_1(\s) =r_2(\s) =  \frac{Q}{\s} \cos \a, \quad
r_3(\s) = \frac{Q}{\s}
\ee
and 
\be \label{tilting}
\tan \a = 2 \pi \a' B.
\ee
Note that the ellipsoid is centred at the coordinates $(x^1,x^2,x^3)
=(0,0,  \s \tan \a)$ and describes D1-strings tilted away from the
normal to the 
D3-brane 
by an angle $\a$. 

\subsection{The fuzzy funnel solution of the D1-branes}

One can
also consider the dual description and study the system from the
D1-strings point of view. 
The Born-Infeld action is 
\cite{myers}
\be \label{BI-d1}
S_{D1}=-T_1 \int d^{2}\sigma\, \; 
{\rm STr}\left[\sqrt{-\det\left(
P_{ab} \left[E_{MN}+E_{M I}(Q^{-1}-\delta)^{IJ}E_{JN}\right]+
\l\,F_{ab}\right)\,\det(Q^I{}_J)}
\right],
\ee
where
\be
E_{MN}=G_{MN}+B_{MN},
\qquad{\rm and}\qquad
Q^I{}_J := \delta^I{}_J+i\lambda\,[\Phi^I,\Phi^K]\,E_{KJ}
\ee
and $\Phi^i$ are scalar fields of mass dimension. The matrix transverse 
coordinates are defined by $X^i = 2 \pi \a' \Phi^i$. Here
$a,b=\s,\t$ are the worldvolume indices of the D1-branes, 
$I,J,K = 1,\cdots,8$
are the indices of the transverse space, and $M,N$ are the 
ten-dimensional spacetime indices.
As was shown in \cite{myers1,myers2},
the static solution of the non-Abelian Born-Infeld theory of the
D1-branes satisfies the Nahm equation
\be \label{nahm}
\del_\s \Phi_i =  i \frac{1}{2} \e_{ijk} [\Phi^j,\Phi^k],
\ee
where $(\t,\s)$ 
are the worldsheet coordinates of the D1-branes
and $\e_{123} =1$. 
The solution $\Phi =0$ corresponds to an infinitely long 
bundle of coincident D1-branes. 
In \cite{myers1}, another solution was found by allowing a 
singular boundary condition\footnote{
This singular boundary condition was first discussed in the context of
D-branes in \cite{d1}.  Some other boundary conditions were considered
in \cite{d2}. 
We emphasis that these are a different kind of boundary condition from
what we are going to derive in the next section.
} 
at $\s=0$ (see \eq{f} below). 
This new solution describes a
fuzzy 2-sphere
\be
\Phi^i(\s) = f(\s) \a^i,
\ee
where $\a^i$ form an $N\times N$ representation of the generators of
an $SU(2)$ subgroup of $SU(N)$, $[\a^i ,\a^j] = 2 \e^{ijk} \a^k$ and
$f$ is given by
\be \label{f}
f =  
\frac{1}{2 \s}. 
\ee
Note that $\Phi$ diverges at $\s=0$. As emphasised in \cite{myers1}, this
new feature was essential to their construction.

This fuzzy funnel solution  carries the 
correct RR charge and tension to be identified with an orthogonal
D3-brane. It also  
matches nicely with the BIon solution in the large $N$ limit. 
To see this, we note that at a 
fixed point $\s$ on the D1-branes, the geometry is that of a fuzzy
sphere with radius $R$ given by
\be
R^2 = \frac{1}{N}\tr (X^i)^2.
\ee 
This gives for the above solution 
\be \label{r-sigma-2}
R(\s) =  \frac{\pi \a' \sqrt{N^2-1}}{|\s|} \approxeq 
\frac{\pi \a' N}{|\s|}
\ee
for large $N$.
This matches precisely with the relation \eq{bion-0}
derived above for the BIon solution.

One may also add a NS $B$-field in the spatial directions transverse to
the D1-brane. For a constant $B$ field with $B_{12}:= B \neq 0$, 
the  effect of the $B$-field to  the D1-brane Born-Infeld action 
was found to modify the Nahm equation \cite{KC} to
\be \label{nahm-B}
\del_\s \phi^i =   i (\frac{1}{2} \e_{ijk} [\phi^j,\phi^k] + \d^3_i i
B),
\ee
where the rescaled fields $\phi^i$ are defined by 
\be \label{scaling1}
\phi^1 := \sqrt{1+(2\pi \a' B)^2} \Phi^1, \quad  
\phi^2 := \sqrt{1+(2\pi \a' B)^2} \Phi^2, \quad
\phi^3 :=\Phi^3.
\ee
The modified Nahm equation has the solution 
\be \label{soln-phi}
\phi^i  = f(\s) \a^i 
- \d^i_3\; B\,  \s,
\ee
where $f$ is the same as \eq{f} above.  To
compare with the BIon solution described in section 2.1, one should
go the description in terms of $X^i$. Using the scaling \eq{scaling1},
the solution \eq{soln-phi} becomes a fuzzy ellipsoid with radii 
\eq{radii} and a tilting   given precisely as in \eq{tilting}. 

Before we close this section, it is
instructive to rewrite the Nahm equation \eq{nahm-B} in terms of the
physical variables $X^i$:
\bea \label{nahm-B-X}
\del_\s X^i &=&  \frac{i}{\l} \epsilon^{ij}[X^j, X^3], \quad i,j=1,2, \nn\\
\del_\s X^3 &=&   \frac{i}{\l} 
(1+(2\pi \a' B)^2)\Big( [X^1,X^2] + i \th \Big),
\eea
where $\th$ is the constant
\be \label{theta}
\th :=  \frac{2\pi \a' B}{1+(2\pi \a' B)^2} .
\ee
In the 
following  two sections 
we will show that the (modified) Nahm equation
can be derived as a 
boundary condition for the open 
D1-strings or F1-strings ending on a D3-brane. 
In particular the understanding of the Nahm equation as boundary
condition of the F1-strings
will provide us 
a understanding to the modification of the Nahm equation 
\eq{nahm-B-X} 
in the presence of a constant 
$B$-field in terms of the noncommutative geometry 
of the D3-brane. 

\section{Nahm Equation as Boundary Condition of D1-strings}

In this section we 
show how to
derive the Nahm equation as a boundary condition of the matrix theory of
D1-strings. 
Let us start by writing down the action of the matrix theory of
D1-strings. This can be derived by taking the Yang-Mills ($\a' \to 0$)
limit of the Born-Infeld action \eq{BI-d1} for the D1-strings.
%
The fluctuations give the matrix string action (the bosonic part)
\be \label{mat-d1}
S_{D1} =
-\frac{1}{\l  g_s}
\int d^2 \s \tr \left(
\frac{1}{2} (\del X^I)^2 -\frac{1}{4 \l^2} [X^I,X^J]^2 +
\frac{\l^2}{4}F_{ab}^2
\right),
\ee
where $X^I = \l \Phi^I$ and $I,J =1,\cdots, 8$.
Since D1-strings are magnetic sources for the Abelian gauge field on
the D3-brane, they couple to the dual Abelian gauge field 
through the boundary coupling
\be \label{gsA}
\frac{1}{g_s}
\int d\t\;  \tr \left[\tilde{A}_\m (x^\n(\t)) \frac{d X^\m(\t)}{d \t} \right],
\ee
where
\be
x^\n(\t,\s) := \frac{1}{N} \tr X^\n(\t,\s)
\ee
is the center of mass coordinate of the matrix string coordinates and
the sum is over $\m=0,1, \cdots 3$. We will comment more on this coupling
in section \ref{AXBdryCoupling}. 
Here $\tilde{F} := \mathrm{d}\tilde{A}$
is the Hodge dual of the field strength $F:= dA$. In general, under
$S$-duality, $F$ is transformed to $S[F]$, which in the leading order
is $S[F] = \Ft$.

With this coupling, 
the boundary conditions for the fields $X^I$ of the D1-strings are 
($\s = 0$):
\bea
X^{i'} &=& \mbox{fixed}, \qquad \qquad  i' =4, \cdots, 8, \label{d1-bc1} \\
\frac{1}{2\pi \a'} 
\del_\s X^i &=& \Ft^{0i} + \Ft_j{}^i \del_\t X^j,  \quad 
 \quad \, i =1, 2, 3,
\label{d1-bc2}
\eea
where $\Ft$ is the field strength of the gauge field $\At$. \eq{d1-bc2}
can be written in a more illuminating way as
\be
\frac{1}{2\pi \a'} (\del_\s X^i+ 2 \pi \a' C^i{}_j \del_\t X^j)  =\Ft^{0i},
\ee
where  $C_{ij} :=  \Ft_{ij}$ corresponds to a RR 2-form potential
and the LHS is simply the mixed boundary condition one would obtain
for the D1-strings if there is a RR 2-form. 
Since there is only a NS $B$-field in our background, 
we will consider the case that $\Ft_{ij} =0$ and the boundary condition
\eq{d1-bc2} reduces to 
\be
\frac{1}{2\pi \a'} \del_\s X^i   =\Ft^{0i}.
\ee
Note that the electric field $\Ft^{0i}$ is by definition equal to the
magnetic field
$B_i = \frac{1}{2} \e_{ijk} F_{jk}$. 
With  the coordinate dependence explicitly spelt out, the boundary
condition \eq{d1-bc2} for the 
D1-strings 
can thus be written as
\be \label{d1-bc3}
\frac{1}{2\pi \a'} 
\del_\s X^i(\s=0) = \frac{1}{2} 
\e_{ijk} F^{jk} (x^1, x^2,x^3)
\quad \, i,j,k =1, 2, 3,
\ee
where $x^i =\tr(X^i(\s=0)) /N$ are the coordinates of the D1-strings endpoint on
the D3-brane. 
This is the equation we are interested in. 
Since this equation tells us about the 
presence of the D3-brane and its properties,
one is tempted to  think of \eq{d1-bc3}
as the defining equation for the D3-brane. Note that since
$F_{ij}$ is a singlet in the $U(N)$ gauge group of the 
D1-strings 
theory, 
$X^i(\s=0) = x^i \id$ and \eq{d1-bc3} becomes a simple
condition on the boundary coordinates of the center of mass string. 

\psfrag{E}{endpoint $\s=0$}
\psfrag{D3}{$D3$}
\psfrag{X123}{$=\{1,2,3\}$}
\psfrag{X9}{$X^9 =\s $}
\psfrag{s0}{matching point $X^9=\s_0$}
\psfrag{D1}{$D1$}
\psfrag{D1X}{$D1$ with excited configuration in $X$}
\psfrag{flat}{}
\psfrag{a}{(a)}
\psfrag{b}{(b)}
\psfrag{B}{$B$}
\EPSFIGURE{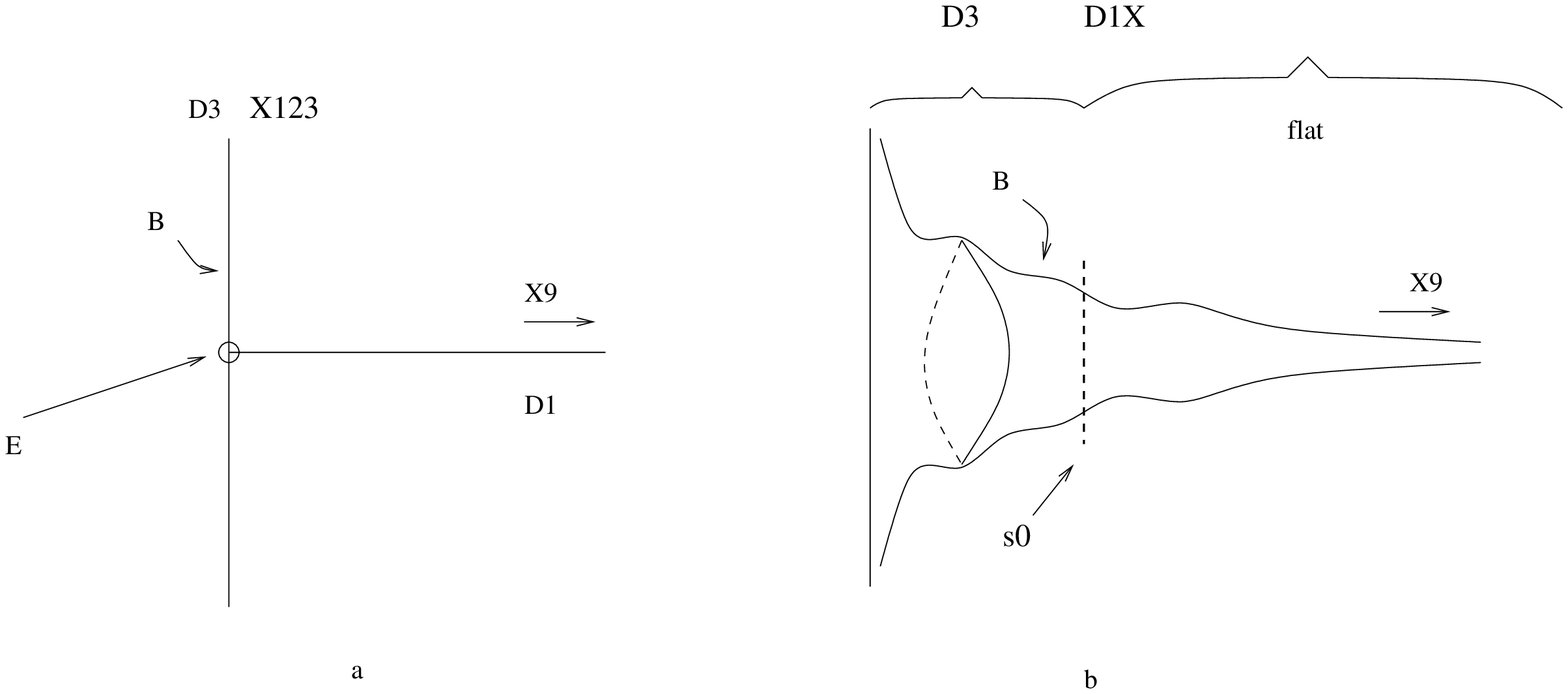,width=15cm}
{Boundary  condition of D1-strings ending on D3-brane. }

However, from the very general spirit of matrix model, one would expect to 
be able to construct the higher dimensional D3-brane from the 
D1-strings
by exciting the transverse scalar fields of the 
D1-strings theory. 
In Figure 1a, the circle denotes the endpoint of the open string where 
the equation  \eq{d1-bc3} holds. By exciting the transverse scalars on
the 
D1-strings, the D3-D1 system becomes something like in Figure 1b. 
It is somewhat arbitrary whether 
one would like to interpret the
transverse scalars $X^i$  as describing the geometry (in a generalized sense) 
of the {\it D3-brane} or as a scalar field configuration of the {\it
D1-branes}
theory. Let us consider a dividing point at $X^9=\s_0 >0$ and interpret the 
region on the LHS of it as describing a D3-brane, and the region on the
RHS as describing 
the D1-strings, 
then the boundary condition \eq{d1-bc3}
becomes
\be \label{d1-bc4}
\frac{1}{2\pi \a'} 
\del_\s X^i(\s_0) = \frac{1}{2} 
\e_{ijk} F^{jk} (X^1, X^2,X^3),
\ee
where $X^i =X^i(\s_0)$ denotes the matrix coordinates of the 
D3-brane and describes the "geometry" of the D3-brane. At the same time
$F_{ij}$ also becomes an $N\times N$ matrix.

Perhaps the simplest way to interpret \eq{d1-bc4} is to consider $N$ widely
separated D1-strings. Each would have a boundary condition of the form of
\eq{d1-bc3}
\be \label{Single-d1-bc3}
\frac{1}{2\pi \a'}
\del_\s X^i = \frac{1}{2}
\e_{ijk} F^{jk} (x^1, x^2,x^3)
\quad \, i,j,k =1, 2, 3,
\ee
where $X^i$ are the coordinates of the D1-string and $x^i = X^i(\s=0)$ are
the coordinates of the D1-string endpoint on the D3-brane. Now, in general we
have an $N \times N$ matrix description of the D1-strings coordinates, and we
would expect to recover these $N$ individual boundary conditions only for
configurations where we could simultaneously diagonalise the matrices.
Therefore, in general we expect a matrix equation of the form
\be \label{N-d1-bc3}
\frac{1}{2\pi \a'}
\del_\s X^i = \frac{1}{2}
\e_{ijk} F^{jk}
\quad \, i,j,k =1, 2, 3,
\ee
where $F^{ij}$ is an $N \times N$ matrix encoding the value of the D3-brane
U(1) field strength at the endpoints of the $N$ D1-strings, whose positions are
given by the eigenvalues of the $N \times N$ matrices $X^i$. The matrices
$F^{ij}$ can be viewed as arising from a standard matrix discretisation of two
worldvolume coordinates of the D3-branes, transforming the functional dependence
of the fields on these coordinates into the discrete labels of the resulting
matrices. When we further
consider the situation at large $N$ illustrated in Figure 1b where the
D1-strings actually generate the D3-brane, the D3-brane U(1) field strength,
and hence the $N \times N$ matrix $F^{ij}$ in \eq{N-d1-bc3} must be
constructed from the fields in the D1-strings action. We therefore arrive at
\eq{d1-bc4}. Note that this construction of the D3-brane could alternatively be
viewed as a specific discretisation of the D3-brane.

Now, the precise form of 
$F^{ij}(X)$ can be fixed by requiring that the action of a
configuration supported by $X^i$ in the matrix
string theory should match with that of the corresponding YM terms of
the D3-brane action
\be \label{mat-d3}
S_{D3} 
=
- T_3  \int d^4 x \left(
 \frac{1}{2}(\del_\m X^A)^2  + \frac{\l^2}{4} F_{\m\n}^2 
\right)
\ee
in the large $N$ limit. 
Doing so requires that 
\be \label{iden-F}
F^{ij} = \frac{i}{(2\pi \a')^2} [X^i, X^j], \quad i,j =1,2,3
\ee
and
\be \label{iden-d4x}
 \int d^4 x = (2\pi l_s)^2 \int d^2 \s\,  \tr 
\ee 
where we have used $T_1/T_3 = (2\pi l_s)^2$.

With this identification, the boundary condition for the matrix  
theory of D1-strings reads:
\be \label{d1-bc}
\del_\s \Phi^i =   
\frac{i}{2} \e_{ijk}[\Phi^j, \Phi^k],
\ee 
at $\s = \s_0$.
However since the dividing point $\s_0$ is 
completely arbitrary, we conclude that the
D1-D3 system depicted in figure 1b can be described by \eq{d1-bc} for
arbitrary $\s \geq 0$. 
Amazingly this is precisely the Nahm equation \eq{nahm} derived from
the Born-Infeld theory of the 
D1-strings. 
Here we have derived it as a
boundary condition of the 
D1-strings 
matrix model.

We remark that 
the identification \eq{iden-F} is derived in the large $N$ limit where
the descriptions of the D1-D3 system in terms of the D3-brane
Born-Infeld theory and the D1-branes Born-Infeld theory completely
overlap \cite{myers1,myers2}. For finite $N$, there will be higher
derivative corrections to the Born-Infeld actions. In general the
two descriptions overlap only for a range of $\s$ which depends on
$N$ and the
identification \eq{iden-F} holds in this range. Outside this range,
the identification \eq{iden-F} and equation \eq{d1-bc} 
will receive $1/N$ corrections.

One can repeat the above analysis to include a $B$-field in the
transverse directions of the D1-strings. Given the non-Abelian
Born-Infeld action \eq{BI-d1} for the D1-strings, one can
derive how the $B$-field modifies the matrix string action. We obtain the
following relevant terms in the D1-strings action:
\bea \label{matrix-d1-B}
S_{D1} =  -\frac{1}{\l  g_s} \int  &&\frac{d^2 \s}{\sqrt{1+\l^2 B^2}}
\tr \left[
\frac{1}{2} (\del \phi^i
)^2
  \right.\nn\\
&&  \left. \qquad -\frac{\l^2}{2} \Big( [\phi^1,\phi^3]^2 + [\phi^2,\phi^3]^2 +
([\phi^1,\phi^2]+ iB)^2 \Big) + \cdots
\right] . \;\;\;
\eea
Identifying
the scalar interaction terms in the D1-strings action with $F^2$ terms
in the D3-brane YM action derived from \eq{BI-d3} with $B:= B_{12} \neq
0$:
\be
S_{D3} = - T_3 \int \frac{d^4 x}{\sqrt{1+\l^2 B^2}} \left[
\frac{\l^2}{2} \Big( F_{13}^2 + F_{23}^2 + F_{12}^2 \Big) + \cdots
\right],
\ee
we need \eq{iden-d4x} and
\be
\label{mod-iden-F}
F^{ij} = i [\phi^i,\phi^j] -B \e^{3 ij}, 
\ee
which is 
a modified version of the identification
\eq{iden-F}. 
As we will argue below, the 
boundary condition in the presence of the $B$-field should take the form
\be \label{bc-d11}
\del_\s \phi^i = \frac{1}{2} \e_{ijk} F^{jk}. 
\ee
Therefore we obtain 
precisely the modified Nahm equation \eq{nahm-B} when
\eq{mod-iden-F} is substituted into  \eq{bc-d11}.
This is how the Nahm equation gets modified by the $B$-field from the point
of view of the D1 matrix string theory.

We note that the boundary condition \eq{bc-d11} 
can be obtained from \eq{matrix-d1-B} by including the boundary 
coupling 
\be \label{gsA-B}
\frac{1}{g_s\sqrt{1+\l^2 B^2}}
\int d\t\;  \tr \left[\tilde{A}_i (x^\n(\t)) \frac{d \phi^i(\t)}{d \t} \right]
\ee
in the gauge $\At_0 =0$.  
This coupling can be seen by the following argument.  
We note that  \eq{matrix-d1-B} has an $SO(3)$
invariance if we transform the fields $\phi^i$ and the external field
$B_{ij}$ as a triplet of $SO(3)$. The boundary coupling should
therefore respect the same symmetry. This is so because, 
due to the tensor gauge symmetry
which keeps $B_{ij} + F_{ij}$ invariant, one can see that  
$\e_{ijk} \Ft^{0k}$, i.e. $\At^i$ transforms in the same way as 
a triplet of $SO(3)$. As for the overall factor in \eq{gsA-B}, we note
that a NS $B$-field with only $B_{12} \neq 0$
should affect the magnetically charged D1 endpoint only in the 3rd
direction. This fixes the overall coefficient since $\phi^{1,2} =
\sqrt{1+\l^2 B^2} X^{1,2}$.

We remark that 
a similar analysis cannot be carried through
to the M2-M5 system since a
Born-Infeld type action for multiple M2-branes which includes the effect
of the $C$-field is not available
\footnote{
See \cite{m2-BI} for some 
results on nonlinear generalization of the BL theory, 
but note that these results do not contain a $C$-field.
}.
Therefore to derive the $C$-field modified Basu-Harvey equation as a
boundary condition of M2-branes, we will have to follow a different route. 
 
In the next section we will provide another new way to understand 
the Nahm equation: as  boundary condition of the
system of F1-strings probing the D1-D3 system. Technically this approach
has the advantage that  the coupling
of the NS $B$-field to the 
F1-strings 
is much simpler than that to the
D1-strings.
Conceptually, this approach allows us to understand this
shift in the Nahm equation as the noncommutative geometry of the
D3-brane in the presence of 
a $B$-field. We will see that this approach 
can be
applied directly to the M2-M5 system and allows us to derive the form
of the quantum geometry over the M5-brane in the presence of
a $C$-field.

\section{Nahm Equation from Boundary Condition of F1-strings}

\subsection{Matrix theory of IIB strings}

According to the proposal of Banks, Fischler, Shenker and Susskind
\cite{BFSS}, the full M-theory in the infinite momentum frame is
described by the large $N$ limit of a $U(N)$ supersymmetric quantum
mechanics of D0-particles where $N$ 
is the number of D0-particles.
Compactification of the matrix theory on a spatial circle yields a
description of the IIA superstrings by a two-dimensional supersymmetric
Yang-Mills theory \cite{motl,BS,dvv}. In particular, by performing a
9-11 flip,  Dijkgraaf, Verlinde and Verlinde 
\cite{dvv} obtained  a description of the  
type IIA string theory in the lightcone gauge in terms of 
a $(1+1)$-dimensional $U(N)$ Yang-Mills theory with $\cN=8$
supersymmetry 
\bea 
S= \frac{1}{2\pi \a'} \int \tr \Big(
\frac{1}{2} (D_\m X^I)^2 &+& \pi^2 g_s^2 l_s^4 F_{\m\n}^2 
-\frac{1}{4\pi^2g_s^2l_s^4}[X^I,X^J]^2  \nn  \\
 & +&  \th^T \Ds \th + \frac{1}{2\pi g_s l_s^2}  \th^T \G_I [X^I, \th]
\Big) . \label{dvv}
\eea
Here the 8 scalar fields $X^I$, $I=1\cdots, 8$ and the 8 fermions
fields $\th^\a_L$, $\th^{\da}_R$  are $N\times N$ Hermitian matrices. 
The fields $X^I, \th^\a, \th^{\da}$ transform respectively 
in the ${\bf 8}_v$ vector, and ${\bf 8}_S$ and ${\bf 8}_c$ spinor
representations of the $SO(8)$ R-symmetry group.  

As usual, the spacetime
coordinates of the strings are identified with the eigenvalues of the $X^I$.
The integral is over a
cylindrical $(1+1)$-dimensional space, with coordinates $\tau$ and
$\sigma \sim \sigma + 2\pi$. The fields do not necessarily have the same
periodicity since they can be identified up to a gauge transformation under
this shift in $\sigma$. Such a gauge transformation can permute the eigenvalues,
leading to the interpretation of the configuration as a collection of closed
strings, each with some periodicity $\sigma \sim \sigma + 2\pi n$, where this
integer $n$ also determines the lightcone momentum of this string. Hence the
integer $N$ determines the total lightcone momentum, not the number of closed
strings.

The Yang-Mills
coupling is dimensionful and is related to the string coupling and the
string length as
\be
\gym= \frac{1}{g_s l_s}.
\ee
It was shown \cite{dvv} that in the IR limit of the SYM theory, the
theory become strongly coupled and the diagonal components of the
$N\times N$ matrices decouple from the off-diagonal components in this
$g_s \to 0$ limit. Moreover, it was shown that 
the IR dynamics of the diagonal elements 
is
described by a sigma model with orbifold target space
$(R^8)^N/S_N$ and the Hilbert space of quantum states of this conformal
field theory is  the same as those for the second quantized type IIA 
superstring theory. 

A T-duality relates the theory of IIA strings with the theory of IIB
strings, where the same action (modulo a different chirality assignment
of the spinors) now describes IIB strings. 
In this description, the action \eq{dvv} with
the spinors $\th^\a, \th^{\da}$ being both ${\bf 8}_S$ of $SO(8)$
describes IIB strings stretched in the 9-th direction. 
Another way to derive this
result is to note that under T-duality (i.e. instead of doing the 9-11
flip), the compactified BFSS matrix theory describes a theory of
D1-strings. A further S-duality turns it 
into the theory of IIB F1-strings.
Note that $N$
measures the lightcone momentum $p_+$ in the IIA matrix string theory,
while it measures the winding number in the IIB matrix string picture.

\subsection{Matrix open string coupling to non-Abelian gauge field}
\label{AXBdryCoupling}

Consider a stack of $N_p$ D$p$-branes whose worldvolume theory is given by
a $U(N_p)$ gauge theory. We would like to derive the coupling of the
worldvolume gauge field $A_\m$ to the matrix open string. Denote the
worldvolume coordinates by $x^\m$, $\m =0,1,\cdots, p$. The gauge field
transforms as
\be
A_\m \to U^{-1} A_\m U + U^{-1} \del_\m U. 
\ee 
We propose a coupling of the form naturally expected for a single string ending
on a single D$p$-brane
\be \label{gcp1}
\int d\t\;  \tr \left[A_\m (x^\n(\t)) \frac{d X^\m(\t)}{d \t} \right],
\ee
where
\be
x^\n(\t,\s) := \frac{1}{N} \tr X^\n(\t,\s) 
\ee
is  the center of mass coordinate of the matrix string coordinates and
the sum in \eq{gcp1} is over $\m=0, 1, \cdots p$.

In the following we make a few comments about this coupling.


{\bf 1.} We note that the above coupling can be made gauge covariant 
in the
following manner. For convenience, we introduce a noncommutativity over the
worldvolume
\be
[y^\m,y^\n] = i \th^{\m\n}, \quad \mbox{$\th^{\m\n}$ constant and
invertible},
\ee
and introduce the field
\be
C_\m := - i (\th^{-1} y)_\m + A_\m(y). 
\ee
Unlike $A_\m$,  $C_\m$ transforms covariantly
\be
C_\m \to U^{-1} C_\m U. 
\ee
Note that since $[C_\m,C_\n] = i \th^{-1}_{\m\n} + F_{\m\n}$, where $F_{\m\n}
:= \del_\m A_\n - \del_\n A_\m + [ A_\m,A_\n]$ is the field strength for
$A_\m$,
\be
\tr [C_\m,C_\n]^2 = N \th^{-2}_{\m\n}  + \tr F_{\m\n}^2.
\ee
The first term on the RHS is a constant and can be dropped. 
Therefore we can use $\tr [C_\m,C_\n]^2$ as the Lagrangian 
and the description in terms of $C$ has a  smooth limit as $\th \to 0$. 
The coupling \eq{gcp1} can be made gauge invariant  by
replacing $A_i$ with $C_i$ and $ d X^i/d \t$ by $D_\t X^i$. However, it
is the term above which describes the dynamics of the open strings on the
D-branes.

{\bf 2.} If we consider precisely the above coupling, there is an
obvious problem. In general $N_p \ne N$ so the matrix multiplication and
trace don't make sense. However, this is not an issue in the special
cases where either $N=1$ or $N_p=1$, and it is this latter case which is
of interest to us. However, for completeness we comment on the general
case. First, note that we also need appropriate boundary conditions for
the (matrix) coordinates $X^{i'}$ orthogonal to the D$p$-branes, i.e.\
strings should actually end on the D$p$-branes. This condition should be
implemented by requiring the eigenvalues of the $X^{i'}$ at $\s=0$ to be
fixed to values corresponding to the position of a D$p$-brane (which
itself is given by the eigenvalues of scalars in the non-Abelian theory
of the $N_p$ D$p$-branes.) Intuitively, this is simply the requirement
that each string ends on a specific D$p$-brane. How many strings end on
each D$p$-brane is given by a choice of boundary conditions for the
system. However, this mapping of $N$ string positions to $N_p$
D$p$-brane positions, interpreted as a mapping between the $N \times N$
and $N_p \times N_p$ matrices describing the string and D$p$-brane
positions, will also provide a way to match the $N \times N$ matrices
describing the string positions $X^i$ parallel to the D$p$-branes with
the $N_p \times N_p$ matrices describing the gauge field on the
D$p$-branes. However, as stated above we will only explicitly consider
the case $N_p=1$ where the matching is trivial.

{\bf 3.} Finally, we note that the coupling of a single open string to
$N_p$ D$p$-branes is usually described by inserting a Wilson loop into
the standard open string partition function, rather than simply
including a boundary coupling in the open string action as we have done
here. However, as discussed in \cite{szabo} these two approaches are
related. In fact the description of a single open string ending on a
stack on $N_p$ D$p$-branes in terms of a Wilson loop is expected to
arise in the effective action, after integrating out interactions
between the $N_p$ D$p$-branes, giving rise to a ``fat brane'' with an
open string ending on it.

\subsection{Nahm equation and noncommutative geometry of the D3-brane}
 
Let us now use the IIB matrix string theory as a probe to study 
the D1-D3 system and to derive the Nahm equation (see Figure 2).  
Since one can identify the endpoint of the open F1-string 
with the D3-brane worldvolume, so one should be able to
understand the Nahm equation as a boundary condition of the open
F1-string. In order to identify the worldvolume coordinates of the
D3-brane with the F1-string, we need to consider a matrix theory of
F1-string based on $U(N)$ matrices.

As we have seen in 
the last section, a NS $B$ field will modify the (bulk)
scalar interaction of the D1-strings matrix theory quite complicatedly. 
On the other hand, the $B$-coupling is a much simpler minimal coupling
for the F1-string since it is an electric source for the $B$-field.  
For matrix strings, 
the interaction with the $B$-field is naturally 
incorporated by the generalization of the usual
coupling 
\be
\label{F1-B-coupling}
S_B := \frac{1}{4\pi \a'}
\int  2 \pi \a' B_{\m\n} \; \tr  D_a X^\m D_b X^\n \e^{ab}. 
\ee
With a constant
$B$-field, the boundary condition
of the matrix F1-string is modified to
\be \label{mixed-bc-source0}
\frac{1}{2\pi \a'} (\del_\s X^i +  2 \pi \a' B^i{}_j \del_\t X^j )
= F^{0i} + F_j{}^i \del_\t X^j 
\ee
at the endpoint $\s =\s_0$.  It is convenient to group the last term
with the $B$-term on the LHS and write
\be
\frac{1}{2\pi \a'} \left(\del_\s X^i +  2 \pi \a' (F^i{}_j + B^i{}_j) 
\del_\t X^j \right)
= F^{0i}.
\ee
Note that the LHS is precisely the boundary condition for a F1-string in
the presence of a background $F+B$. Since we would like to describe a
D3-brane with a constant $B$-field on it but with no background $F_{ij}$,
we will let $F_{ij} =0$ and the boundary condition becomes
\be \label{mixed-bc-source1}
\frac{1}{2\pi \a'} \left(\del_\s X^i +  2 \pi \a' B^i{}_j 
\del_\t X^j \right)
= F^{0i}.
\ee
Let us introduce the dual field strength $\Ft$ defined by:
\be
F^{\m\n} = \frac{\sqrt{-G}}{2} \e^{\m\n\a\b} \Ft_{\a\b} 
= \frac{1}{2 \sqrt{-G}} \e_{\m\n\a\b} \Ft_{\a\b}.
\ee
Here the convention for the  Hodge duality operation is
$\e_{0123} =1 =-G \e^{0123}$. 
The boundary condition for the F1-strings thus takes the form,
\be\label{mixed-bc-source}
\frac{1}{2\pi \a'} (\del_\s X^i +  2 \pi \a' B^i{}_j \del_\t X^j )
=  \frac{1}{2}\frac{\e_{ijk}}{\sqrt{-G}} \Ft_{jk},
\ee
For reasons that will become clear later,
we have allowed for the possibility that a nontrivial metric
$G_{\m\n}$ may appear on the D3-brane worldvolume when $B\neq 0$. 

To proceed, we need to identify $F^{0i}$ with a 
configuration of the boundary  values of the 
scalar fields $X^i$ of the F1-strings.  We {\it propose}
that
\be \label{iden-F-univ}
\Ft^{ij} = \frac{i }{(2\pi \a')^2} [X^i,X^j], \quad i,j =1,2,3,
\ee
This is the S-dual of the relation \eq{iden-F}. 
This is reasonable as the $B$-field couples to
the F1-strings only through the term \eq{F1-B-coupling} and there is
no modification to the bulk scalar interaction as in \eq{matrix-d1-B} 
for the case of 
D1-strings
in a $B$-field. Since the effect of the
$B$-field has already been taken into account by \eq{F1-B-coupling}, 
it is 
reasonable to propose the identification \eq{iden-F-univ} from the
S-dual statement of \eq{iden-F} which holds for 
D1-branes without $B$-field.

When there is no $B$-field, $G_{\m\n} =\eta_{\m\n}$. 
On substituting \eq{iden-F-univ}, the boundary condition 
\eq{mixed-bc-source} yields  
\be
\frac{1}{2\pi \a'} \del_\s X^i =  \frac{i}{(2\pi \a')^2} [X^i,X^j], 
\quad i,j =1,2,3,
\ee
at $\s =\s_0$. 
Here $X^i$ are the matrix coordinates of the F1-strings.  Since the
boundary coordinates can be identified with the 
D1-branes 
coordinates,
and since the F1-strings can end anywhere on the 
D1-branes 
(see Figure 2), 
this equation actually describes the D3-brane
and we arrive precisely at the original Nahm equation \eq{nahm}.

\psfrag{D3}{$D1$ expands as a $D3$-brane}
\psfrag{X9}{$X^9 =\s $}
\psfrag{F1}{$F1$ string probe}
\psfrag{B}{$B_{12} \neq 0$}
\psfrag{s0}{$\s_0$}
\EPSFIGURE{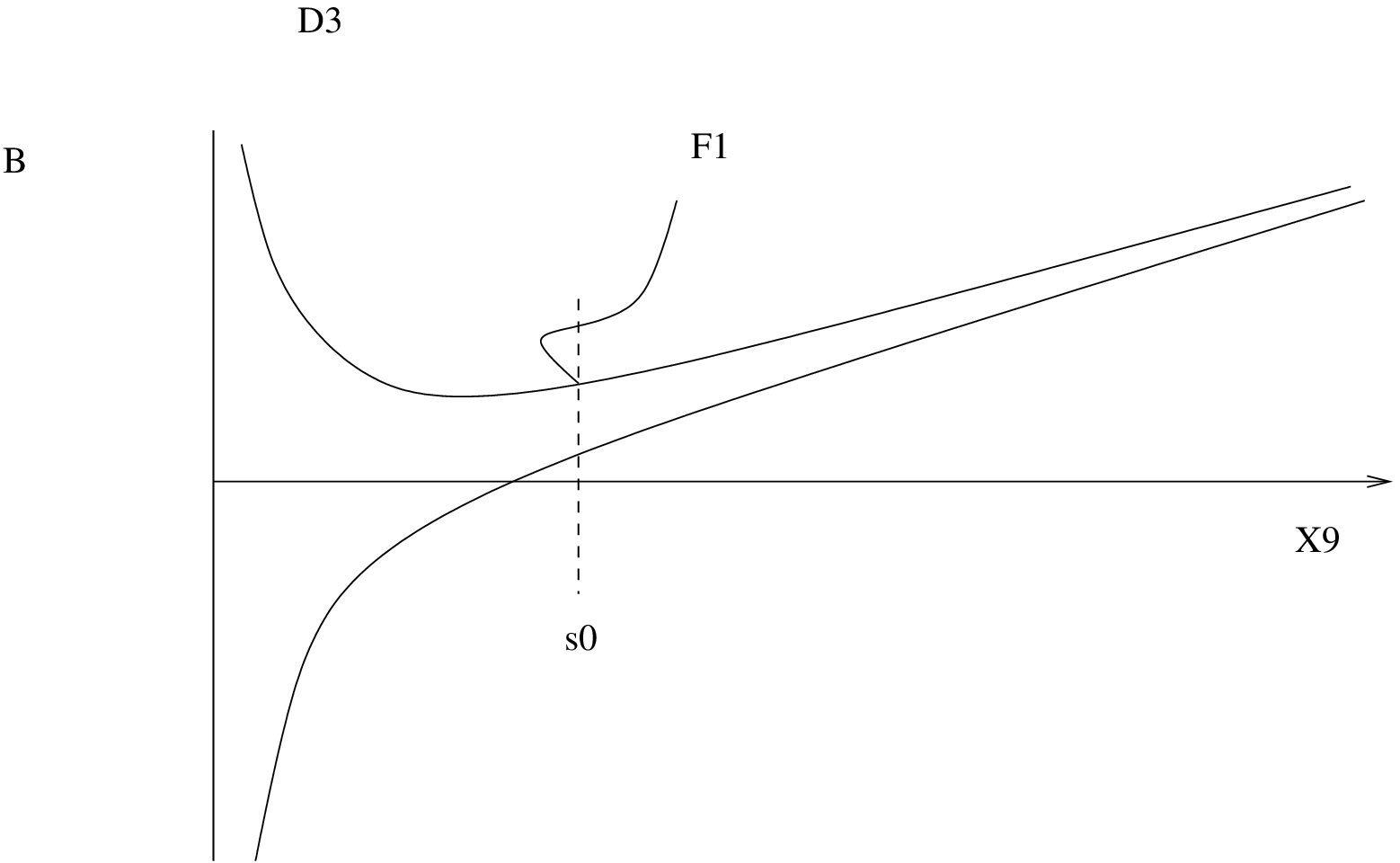,width=10cm}
{ F1-strings probing the D3-brane.  }
 
Next we consider the case with $B \neq 0$. Since turning on $B_{12}$
breaks the Lorentz group $SO(1,3)$  to $SO(1,1) \times SO(2)$, a metric
$G_{\m\n}$ of the following form is allowed:
\be \label{ruler}
G_{\m\n} = \left( 
\begin{array}{cccc}
-g_0 &0&0&0 \\
0&g_1 &0&0 \\
0&0& g_2&0\\
0&0&0&g_3
\end{array}
\right),
\ee 
where $g_0 =g_3$ and $g_1=g_2$. The boundary condition
\eq{mixed-bc-source} reads
\be \label{nnn}
\del_\s X^i +  2 \pi \a' B^i{}_j \del_\t X^j  =
 \frac{i\e_{ijk} }{2\l} \frac{ g_j g_k}{ g_0 g_1} [X^j,X^k] .
\ee
Note that when there is no $B$-field, the equation 
\eq{nnn} can be solved by
a static configuration. With a  $B$-field turned on, this is no longer
the case.  A simple ansatz for $X^i(\t,\s)$ is to make an amendment 
to the static ansatz as follows:
\be \label{a1}
X^i(\t,\s) = X_0^i(\t,\s) \id + Y^i(\s),
\ee
where $X_0^i$ and $Y^i$ satisfy 
\bea  
&& \del_\s X^i +  2 \pi \a' B^i{}_j \del_\t X^j =0, \label{mixed-bc}\\
&& \del_\s Y^i = \frac{i\e_{ijk}  }{2\l}\frac{ g_j g_k}{ g_0 g_1} \left(
[X_0^j, X_0^k]\id + [X_0^j \id, Y^k] + [Y^j, X_0^k\id]+ 
[Y^j,Y^k]\right) \label{a3}
\eea   
at the endpoint of F1-strings.

As was analyzed in \cite{CH1,CH2,chu},  the mixed boundary
condition \eq{mixed-bc} is not compatible with the standard canonical
quantization. The proper quantization taking into account the 
$B$-field 
has been carried out in \cite{CH1,CH2,chu} with the essential result that
the zero modes $x_0^i$ become noncommutative.
These results can be applied here immediately as the quantization
problem is the same. The result in terms of  $X_0^i$ reads 
\be \label{ncg-bdy}
[X^i_0(\t,\s), X^j_0(\t,\s')] =
\begin{cases}
\; i \th^{ij},
 \; & 
\mbox{for $\s  = \s'= \s_0$} \\
\; 0, & \mbox{otherwise}
\end{cases}
\ee  
where $\th^{ij}= (2 \pi \a')^2 \big(B (1-  (2 \pi \a')^2
B^2)^{-1}\big)^{ij}$. For the present case $B=B_{12} \neq 0$, it is 
\be
\th^{12} = \frac{(2 \pi \a')^2 B}{ (1+  (2 \pi \a' B)^2}:= \th ,
\qquad \qquad\mbox{and} \quad \th^{ij}=0, \quad \mbox{otherwise}. 
\ee
The equation \eq{ncg-bdy} is precisely saying that $X^i_0$ is becoming
noncommutative at the point where the F1-strings touch the D1-branes.
Thus $X_0^i$ can be thought of as the  coordinates
describing the underlying noncommutative D3-brane. 

As for equation \eq{a3}, we treat it as a classical equation and so
the commutator of $Y^i$ with $X_0^j$ vanishes. We obtain
\be \label{YY}
\del_\s Y^i = \frac{i}{2 \l} \frac{ g_j g_k}{ g_0 g_1} 
\e_{ijk}\left(
[Y^j,Y^k] + i \th \d^i_3 \right).
\ee 
With $Y^i$ identified with the 
D3-brane 
coordinates, one can think of
\eq{YY}  as the equation describing the 
shape of the D1-branes spike 
in Figure 2.  
Written explicitly, we have
\bea \label{YYY}
\del_\s Y^i &=& \frac{i}{\l} \e^{ij} [Y^j,Y^3], \quad i,j
=1,2, \nn\\
\del_\s Y^3 &=& \frac{i}{\l} \frac{g_1}{g_0} ( [Y^1,Y^2] + i \th).
\eea
These are the equations which describe the D3-brane with $B$-field from
the D1-strings point of view. 

An interesting result of this analysis is that the constant term $\th$ in
\eq{YYY} has a simple interpretation in terms of the noncommutative geometry of
the D3-brane:  $[X^i_0,X^j_0] = i \th^{ij}$.
With this as an input, we can now fix the
metric components  $g_0$ and $g_1$. 
Recall that in the presence of a $B$-field,
the worldvolume theory of a D3-brane is given by a noncommutative
Yang-Mills theory with noncommutativity parameters $\th^{\m\n}$ and with the
open string metric $G_{\m\n}$ \cite{SW}.
In contrast to the closed string metric $g_{\m\n}$, $G_{\m\n}$ is 
the effective  metric seen by the open strings and is the one relevant
for the description of the noncommutative Yang-Mills.
In the present case with only $B_{12} \neq 0$ and $g_{\m\n} =
\eta_{\m\n}$, $G_{\m\n}$ is given by \eq{ruler} with
\be
g_0 =1, \quad g_1 = 1+ \l^2 B^2.
\ee
And our \eq{YYY} reproduces precisely the modified Nahm equation 
\eq{nahm-B-X}.

This understanding of the Nahm equation as boundary conditions for the
F1-strings has provided us with a more precise understanding of the 
physical meaning of the 
$B$-field 
modifications in the Nahm equation
\eq{nahm-B-X} in terms of noncommutative geometry 
($\th^{\m\n}$ and $G_{\m\n}$) of the
open string in a background $B$-field.

\section{$C$-field modification of the Basu-Harvey Equation}

Following a similar analysis as for the Nahm equation, 
we now show that the original 
Basu-Harvey equation describing an M5-brane
can also be understood as a boundary condition of
the theory of multiple membranes. 

One of the interesting features of the theory of the M5-brane is that
the worldvolume field strength $H$ satisfies a nonlinear
self-duality condition.  
In the formulation of \cite{m51}, the three-form field
strength $H_{\m\n\l}$ on the worldvolume of the M5-brane is expressed in
terms of an auxiliary self-dual field strength 
$h_{abc} =\frac{1}{6}\epsilon_{abcdef} h^{def}$ as
\be
H_{\m\n\l}=E^{\ a}_\m E^{\ b}_\n E^{\ c}_\l m^{\ d}_b m^{\ e}_c
h_{ade}=e^{\ a}_\m e^{\ b}_\n e^{\ c}_\l (m^{-1})^{\ e}_ch_{abe}, 
\ee
where
\be
m^{\ b}_a:= \delta^{\ b}_a-2h_{acd}h^{bcd},
\ee
The non-linear self-duality condition for $H$ is 
\be
\frac{\sqrt{-{\rm det}\,g}}{6}\epsilon_{\m\n\l \g \r\s} H^{\g \r\s}=
\frac{{1+K}}{ 2}(\cG^{-1})^{\; \r}_\m H_{\n \l \r},
\label{sd-H}
\ee
where $g_{\m \n}$ is the metric on the M5-brane, 
$K:= \sqrt{1+\frac{1}{24} H_{\m\n\l}g^{\m \g}g^{\n \r}g^{\l \s} 
H_{\g \r \s}}$.
The metric
\be
\label{G2}
\cG_{\m\n}=E^{\ a}_\m E^{\ b}_\n\eta_{ab}
\ee 
has its vielbein  given by 
\be
E^{\ a}_\m:=e^{\ b}_\m (m^{-1})^{\ a}_b, 
\ee
where  $e^{\ a}_\m$ is the vielbein of the metric   $g_{\m\n}$.
It turns out this nonlinear self-duality condition will play an
important role in our analysis below. 

\subsection{The Basu-Harvey equation}

Let us 
first review the Basu-Harvey equation which has been  proposed to
describe the system of multiple M2-branes ending on an
M5-brane. Consider the system with $N$ M2-branes lying in the 0,1,9 directions
and the M5-brane in the 0,1,2,3,4,5 directions. The Basu-Harvey
equation is
\be
\del_2 X^i + i \frac{K}{3!} \e_{ijkl}[X^j,X^k,X^l]=0, \quad i=2,3,4,5 . 
\ee
Here $(\t,\s_1,\s_2)$ are the worldvolume coordinates of the membranes,
$X^i(\s_2)$ describes the transverse fluctuations to the
M2-branes, 
$K$ is a constant of mass dimension 3
and $\e_{2345} =1$. 
In \cite{BH}, $X^i$ is taken to be valued in an algebra  $\cA_4$ with
generators $T^i$, $i=1,\cdots, 4$ and the 3-bracket 
satisfies the $SO(4)$-invariant 
relation
\be \label{fuzzy3}
[T^i,T^j,T^k] = i \e_{ijkl} T^l.
\ee
The Basu-Harvey equation is solved by
\be \label{wedge}
X^i(\s_2) = f(\s_2) T^i, 
\ee
where
\be
f(\s_2) = \frac{1}{\sqrt{2 K \s_2}}.
\ee
Note that the operator $C:=\sum( T^i)^2$ is central in the algebra
\eq{fuzzy3} and so one can consider representations of the algebra
\eq{fuzzy3} with constant values of this operator. For each such
representation, the solution \eq{wedge} describes a wedge of M2-branes
which open up to the M5-brane and whose cross-section is described by
a fuzzy three-sphere
\cite{ram} with radius $r^2(\s_2) := \sum (X^i)^2$ given by
\be
r^2(\s_2) = \frac{C}{2 K \s_2}. 
\ee
By choosing the values of the constants $K$ and $C$ appropriately, one
reproduces precisely the radius-to-distance relation obtained from the
M2-brane solution \cite{HLW} 
solved in the M5-brane field theory
\be
r^2(\s_2) = \frac{Q}{\s_2}.
\ee
where
\be \label{Q-M5}
Q= 3 \pi^2 l_P^3 N
\ee
for the M2-M5 branes system and $l_P=g_s^{1/3} l_s$ is the Planck length. 
For more general solutions to the Basu-Harvey equation and
interpretation in terms of M-branes, see \cite{BC}.

Just as the Nahm equation can be understood as a BPS
equation in the 
D1-branes 
Born-Infeld theory, the Basu-Harvey equation
can also be understood as a BPS equation in the BL theory for multiple
membranes \cite{BLG1,BLG2,BLG3}. 
This was originally derived for the BL theory
based on the 3-algebra $\cA_4$, although the argument applies equally
to a BL theory based on a general Lie 3-algebra. However since the inner
product for a Lie 3-algebra cannot be positive definite apart from the
$\cA_4$ case \cite{metric}, the theory will necessarily contain zero or negative 
norm states. Therefore it is not clear whether these theories, and
the resulting BPS equations, make sense. Attempts have been made to make sense
of the BL theory of multiple membranes 
based on a particular form of Lorentzian 3-algebras \cite{lor}.
The negative norm state there  can be eliminated if one enriches the
theory with additional fields such that there is an additional gauge
symmetry. The resulting theory however turns out to be
equivalent to the D2-brane supersymmetric Yang-Mills theory \cite{d2d2}.
Nevertheless we do not rule out the possibility that for a certain specific 
kind of 3-algebra, the BL theory is physically 
well defined, perhaps after certain extensions and modifications
of the original construction. 

Assuming such a
theory of (an arbitrary number of) 
multiple M2-branes exists,
the Basu-Harvey equation which describes the M5-brane 
where M2-branes end is presumably just the BPS condition of this theory. 
However since we do not have the knowledge 
of the form of the action as well as the  supersymmetry transformations,
we will have to take a different route to derive this equation.
As we will show next, the Basu-Harvey equation can be understood as the 
boundary condition 
of the  M2-branes probing the M2-M5 intersecting system.
The advantage of this approach is that much
less knowledge about the structure of the theory of multiple 
membranes is needed to derive the boundary condition.

\subsection{Basu-Harvey equation as boundary condition of membrane theory}

Since eventually we will be interested in adding a 3-form $C$-field, and
since the Born-Infeld type nonlinear generalization of the
multiple membrane theory including the effects of the $C$-field 
is unknown, let us therefore use  
use the brane-probe
approach to derive the Basu-Harvey equation as the boundary condition 
of the probing  M2-branes.

Let us first consider the case without $C$-field. 
To derive the boundary condition of the membrane, we will assume that
the action (in the static gauge) 
contains the kinetic term  
\be \label{gen-membrane}
S = \int d^3 \s  \frac{1}{2} \langle D_a X^I, D_a X^I \rangle   + \cdots 
\ee
where $\langle \, , \rangle$ denotes an inner product of the 3-algebra
and $\cdots$ denotes fermions and gauge field kinetic terms and interaction 
terms. In the original BL theory, the kinetic term in \eq{gen-membrane}
is the only term that contributes to the boundary condition of $X^I$
when the membranes are taken to be open. The $\cdots$ terms do not contribute.
We will make the mild assumption that this will continue to 
be the case in the general theory of membranes. 
 
Now consider a system of multiple membranes extending in the 0,1,9
directions and 
growing into  
an M5-brane which extends in the 0,1,2,3,4,5
directions. 
For a single membrane ending on an M5-brane, the boundary of the membrane
appears as a non-gravitational string sourcing the self-dual 2-form 
$\cB$
on the M5-brane 
\be
\int \cB_{\m\n} \del_a X^\m \del_b X^\n \e^{ab}, \qquad \m,\n =0,1,\cdots,5,
\quad a,b=\t, \s_1. 
\ee
For multiple membranes, the boundary  becomes
multiple self-dual strings. 
Most generally, one would like to describe the coupling
for multiple self-dual strings to multiple M5-branes, just as the
coupling \eq{gcp1} for multiple strings with multiple D-branes. However
this requires the knowledge of ``non-Abelian'' tensor multiplets and a
representation of the 3-algebra. Without knowing this, we will have to be
content with the coupling of multiple self-dual strings to a single
M5-brane. This is analogous to the coupling of the center of mass string
to a single D-brane $\int   A_\mu \dot{x}^\m$ and can be written down
easily
\be
\int \cB_{\m\n} \langle D_a X^\m, D_b X^\n \rangle \e^{ab}, \quad
a,b = \t, \s_1
\ee
assuming the membrane is
extended in $X^1$ and has a boundary at $\s_2 =0$.
Taking a static gauge $X^0 =\t, X^1=\s_1$, it is easy to derive
the boundary condition for the open membranes. 
It is
\be \label{bc-m2-0}
\del_{\s_2} X^i + F^i{}_{jk}
\del_\t X^j \del_{\s_1} X^k= 
F^{01i} + F^{i0}{}_j \del_{\s_1} X^j + F^{i1}{}_j \del_\t X^j ,
\quad i,j =2,3,4,5, 
\ee 
for $\s_2=0$. 
Here  $F:=d\cB$ and
$H=d\cB + C$
is the self-dual 3-form field strength of the M5-brane.
Note that the LHS is precisely the combination that will appear in the
boundary condition of a membrane in the presence of a background 3-form
$C$-field. In this section we do not consider such a $C$-field, so let
us set the background $F_{ijk} =0$.

As in the case of F1-strings, it is convenient to introduce the  
dual field strength $\Ft$ defined by
\be \label{hd}
F^{\m\n\l} 
:=  \frac{\sqrt{-G}}{3!} \e^{\m\n\l\a\b\g} \Ft_{\a\b\g} 
= \frac{1}{3!\sqrt{-G}} \e_{\m\n\l\a\b\g}\Ft_{\a\b\g} ,
\ee 
where the Hodge star is taken with the convention $\e_{012345} =1 =
-G \e^{012345}$.  We have allowed for the possibility of having a
non-trivial metric $G_{\m\n}$ on the M5-brane in general. $F^{01i}$ can
thus be expressed in terms of the spatial components $\Ft_{jkl}$ 
of the dual field strength.   
As for the components
$F^{0i}{}_j$ and $F^{1i}{}_j$, we will consider the case where
they are equal to zero, which corresponds to a particular configuration
of M5-brane. 
The equation
\eq{bc-m2-0} now reads
\be\label{bc-m2}
\del_2 X^i =  \frac{1}{3!\sqrt{-G}} \e_{ijkl} \Ft_{jkl} , \quad 
i, j,k,l= 2,3,4,5.
\ee

Repeating the probe argument as for the F1-strings, the M5-brane is
defined
by 
the equation \eq{bc-m2} with an appropriate identification of the field
strength $\Ft_{jkl}$ with a certain configuration of the boundary value of 
$X^i$. 
Inserting back properly the dimensional proportional
constant $K$, we propose that 
\be \label{iden-H}
\Ft^{ijk} = i f K [X^i,X^j,X^k], \quad i,j,k =2,3,4,5,
\ee
where, in general, the scalar 
$f$ is a function of the background $C$, and $f=1$ when $C=0$.
We will make more comment on this relation in the 
discussion section. 
For now, let us mention that the relation \eq{iden-H} is consistent
with the relation \eq{iden-F-univ} for the F1-strings.

Back to our analysis.
In the case when there is no $C$-field, 
the equation \eq{bc-m2} reads 
\be \label{gen-BH}
\del_2 X^i = i \frac{K}{3!} \e_{ijkl} [X^j,X^k,X^l].
\ee
when the identification \eq{iden-H} is substituted.
This is precisely the original Basu-Harvey equation.
Here we have derived it as a boundary condition of the open membrane. 
We note that the identification \eq{iden-H}
allows us to represent the worldvolume field strength of the M5-brane in
terms of a 3-bracket of scalar fields of the M2-brane. It will be
interesting to understand better  
its other implications in the physics  of M2-branes and
M5-branes.
We also remark that when there is no $C$-field, 
one can also 
derive the Basu-Harvey 
equation as the boundary condition 
of the original set of M2-branes as in section 3.
 
We also remark that in general it may be possible to find more general
configurations of the M2-branes which also excite the  components 
$F^{0i}{}_j$ and $F^{1i}{}_j$. This will modify the 
equation \eq{bc-m2}.
It would be  interesting to  understand what would be the
identification of these components in terms of $X$ and 
to understand 
the resulting equation which generalizes the Basu-Harvey equation; 
and to study the properties of these more general configurations
of M5-branes.

\subsection{$C$-field modification to 
Basu-Harvey equation and the quantum geometry of the M5-brane}
\label{C_BH}

Let us now incorporate a constant $C$-field on the M5-brane and
ask how it modifies the Basu-Harvey equation \eq{gen-BH}.

To start with, we recall that
the system of M2-branes ending on M5-branes with a constant $C$-field 
($C_{012}, C_{345} \neq 0$) has 
been studied from the M5-brane point of view before \cite{michishita,youm} 
where the M2-brane was constructed as a soliton of the M5-brane equation of
motion. 
In the papers  \cite{michishita,youm}, 
a constant worldvolume field strength $h_{012} = -h_{345}
=h = \frac{1}{2}\tan \frac{\alpha}{2}$ was considered. 
The components of $H$ and the auxiliary field 
$h$ are related by $H_{012} = \frac{\sin \a}{4}$, 
$H_{345} = -\frac{\tan \a}{4}$. 
This is equivalent to 
a background with $F=0$ and a constant 3-form potential
$C$ 
\be \label{C}
C_{012} = \frac{\sin \a}{4}, \quad C_{345} =  -\frac{\tan \a}{4}
\ee
due to the tensor-gauge symmetry which keeps $H=C+F$ invariant.

The M2-brane soliton extends in the $X^9 = \s_2$ direction and 
has a cross-section 
$R\times S_3$ where $S_3$ is an ellipsoid described by
\be \label{ellipsoid-M5}
\frac{x_3^2}{r_3(\s_2)^2} + \frac{x_4^2}{r_4(\s_2)^2} +\frac{x_5^2}{r_5(\s_2)^2} 
+ \frac{(x_2- \s_2 \tan \a)^2}{r_2(\s_2)^2} =1.
\ee
Here the radii are
\be \label{radii-M5}
r_2^2 = \frac{Q}{\s_2\cos \a}, \qquad
r_3^2=r_4^2=r_5^2= \frac{Q}{\s_2} \cos \a = r_2^2 \cos^2 \a, 
\ee
where
$Q$ is given by \eq{Q-M5}.
This solution can be interpreted as a bundle of $N$ M2-branes coming out of 
the M5-brane. The $\s_2$-dependent shift in the $x_2$ tells us 
that the M2-branes wedge is tilted away from
the normal to the M5-brane 
with an angle $\a$.
Similar to the case of the 
D1-branes,
the amount of tilting is determined by a balance of the pull on the
boundary of the M2-branes due to the $C$-field component $C_{012}$ and 
the  M2-brane tension.

To understand these results in terms of the M2-branes, we propose the following
simple modification to the  Basu-Harvey equation:
\be \label{gen-BH-C}
\del_2 \phi^i =  \frac{i \b}{3!} \e_{ijkl} [\phi^j,\phi^k,\phi^l] 
- \d^i_2 K^{1/2} \tan \a,
\ee
where
\be
\b = \cos \a ,
\ee
\be \label{scaling-X}
\frac{\phi^i}{ K^{1/2}}:= \begin{cases} (1+\tan^2 \a)^{1/2} X^i, & 
 \mbox{for $i=3,4,5$}, \\
X^i, &  \mbox{for $i=2$},
\end{cases}
\ee 
and $\tan \a = 4 C$ where $C:=- C_{345}$. 
Here $X^i$ is the variable to be used to
match with the M5-brane soliton result reviewed above. 
The parameter $\b$, the shift in   \eq{gen-BH-C} and the
scaling 
factors 
in \eq{scaling-X} are fixed by matching with 
the radius $r_2$,  the $\s_2$-shift in \eq{ellipsoid-M5}, and  
the ratio of radii in \eq{radii-M5} respectively. To derive this, we
use the fact that the
equation \eq{gen-BH-C} is invariant under $SO(4)$ rotations, implying
that a solution  which is spherical symmetric when written in $\phi$
can be constructed.  This solution, when rewritten in terms of $X$,
features the ellipsoidal cross sectional 
geometry and the $\s_2$-dependent shift as in \eq{ellipsoid-M5}.

Next we would like to derive the generalized Basu-Harvey equation 
\eq{gen-BH-C} 
as a boundary condition of probe open-membranes.
We start with a single membrane ending on 
an M5-brane with a constant worldvolume $C$-field. 
The coupling of the single membrane to the $C$-field is
\be \label{C1}
\int d^3 \s C_{\m\n\l} \del_0 X^\m \del_1 X^\n  \del_2 X^\l . 
\ee
For a 
membrane  with endpoint at 
$\s_2=\s_{20}$, the boundary condition reads
\be\label{bc-m2-H1}
\del_2 X^i +  C^i{}_{jk} \del_0 X^j \del_1 X^k 
=  \frac{1}{3!\sqrt{-G}} \e_{ijkl} \Ht_{jkl} , \quad 
i, j,k,l= 2,3,4,5.
\ee
Since
our $C$-field configuration breaks $SO(1,5) \to SO(1,2) \times SO(3)$,
one can expect the metric $G_{\m\n}$ to be of the form:
\be\label{G-M5}
G_{\m\n} = \left( 
\begin{array}{cccccc}
-g_0 &0&0&0&0&0 \\
0 &g_0&0&0&0&0 \\
0 &0&g_0&0&0&0 \\
0 &0&0&g_1&0&0 \\
0 &0&0&0&g_1&0 \\
0 &0&0&0&0&g_1 \\
\end{array}
\right)
\ee
For the analysis below, 
we do not assume any relation between this metric and 
the  metric $\cG_{\m\n}$ which appears in the self-duality condition of
$H$.

For multiple membranes, the natural generalization to the coupling
\eq{C1} is
\be \label{C2}
S_C = \frac{1}{3!} \int d^3 \s C_{\m\n\l} \tr 
( D_a X^\m, D_b X^\n,  D_c X^\l) \e^{abc}, 
\ee
where $D_a$ $(a=0,1,2)$ is the covariant derivative on the multiple
M2-branes theory. To write down this coupling, we have 
assumed that the 3-algebra $\cA$ is equipped with a map
\be
\tr : \cA \otimes\cA \otimes\cA \to \CC,
\ee
which is completely symmetric and is invariant in the following sense:
\be \label{d-inv}
\tr([\a,\b,f],g,h)+ \tr([f,[\a,\b,g],h)+ \tr(f,g,[\a,\b,h]) =0.
\ee
In terms of generators, the map $\tr$ can be specified by the constants
\be
d^{abc} = \tr(T^a,T^b, T^c). 
\ee 
An explicit construction of the map $\tr$ has been given before in 
\cite{m2-C} for the case that  the 3-algebra is given by 
the Lorentzian 3-algebra \cite{lor}.
Another example is to take $\tr$ equal to the integration $\int d^3
y$ if the (infinite dimensional) 3-algebra is given by the Nambu bracket
$[f,g,h] = \e^{ijk} \del_i f \del_j g \del_k h$
over a 3-manifold with local coordinates $y^{1,2,3}$. Such an
infinite dimensional 3-algebra has been employed in \cite{M2M5} to
construct the M5-brane theory out of the multiple M2-branes theory.
In general
it is an interesting question to
understand which class of 3-algebras admits such a map.
It appears to us
that this is an essential requirement for a theory of multiple M2-branes
since one must be able to incorporate a coupling to the $C$-field and \eq{C2}
is the most natural candidate of such a coupling.
It would be interesting to understand better this mathematical property and
how is could be incorporated into the classification of 3-algebras suitable for
BL theories \cite{3AlgClass}.
It is possible that one needs a representation of $\cA$.

In general, given the map $\tr$, one can construct a 
linear and symmetrical
product $*$ on the 3-algebra 
\be
* : \cA \otimes \cA \to \cA 
\ee
by
\be
f*g = \frac{d^{abc}}{d^{000}} 
f_a g_b T^c, \quad \mbox{where $f=f_a T^a, g= g_b T^b$},
\ee
where $T^0 = \id$ is the identity operator defined by the property
that $[\id, T^b,T^c] =0$ for all $b,c$ and by the normalization
$\langle \id, \id \rangle$ =1. We will assume that $\cA$ has such an 
(unique) identity operator. If we also assume that, just as for Lie algebra, 
$T^0$ does not appear
on the RHS of a 3-bracket, i.e. $f^{abc}{}_0 =0$ for all $a,b,c$, then it
is easy to check that 
the condition \eq{d-inv} is solved by
$d^{0 a b} = d^{000} \d^a_0 \d^b_0$. This implies that $\id * \id = \id$. 
We will consider 3-algebra of this form in the following. 
With the aid of this product, the boundary condition for the 
probe M2-branes thus takes the form ($i,j =2,3,4,5$),
\be \label{bc-m2-H}
\del_2 X^i +  C^i{}_{jk} \del_0 X^j * \del_1 X^k 
= \frac{1}{3! \sqrt{-G} } \e_{ijkl} G_{jj} G_{kk} G_{ll}  
\Ft^{jkl}
\ee
for $\s_2=\s_{20}$. As before we have set the field strength 
$F^i{}_{jk}, F^{i0}{}_j, F^{i1}{}_j =0$ and content ourselves 
with considering an
M5-brane whose shape is described by the components $F^{01i}$ of the
field strength.

Substituting the identification \eq{iden-H} 
into \eq{bc-m2-H}, we arrive at the equation
\be \label{BH-open}
\del_2 X^i +  C_{ijk} \del_0 X^j * \del_1 X^k  
= \frac{i f K}{3! \sqrt{-G} } \e^{ijkl} G_{jj} G_{kk} G_{ll}  
[X^j,X^k,X^l]  , \quad i,j =2,3,4,5. 
\ee
This equation is what one would expect to define an M5-brane from the open
M2-branes theory.
As in the case of the D3-brane,  
the equation \eq{BH-open} reproduces our Basu-Harvey equation
\eq{gen-BH-C} if we substitute the ansatz
\be
X^i (\t,\s_1,\s_2) = X^i_0 (\t,\s_1,\s_2)\id + Y^i(\s_2),
\ee
where $X^i_0(\t,\s_1,\s_2)$ and $Y^i(\s_2)$ satisfy
\be \label{X1}
\del_{2} X^i_0 + C_{ijk} \del_0 X_0^j  \del_1 X_0^k =0, 
\ee
\be \label{X2}
\del_2 Y^i = \frac{i f K}{3! \sqrt{-G} } \e^{ijkl} G_{jj} G_{kk} G_{ll}
\left( [X_0^j,X_0^k,X_0^l] + [Y^j,Y^k,Y^l]\right) 
\ee
at the endpoint of the probe M2-branes.
Written more explicitly, this reads
\bea
\del_2 Y^2 = i f K (\frac{ g_1}{g_0})^{3/2} &&\left(
[X_0^3,X_0^4,X_0^5] +[Y^3,Y^4,Y^5]
\right), \\
\del_2 Y^i = i f K (\frac{ g_1}{g_0})^{1/2}  \e^{i2jk} && \left(
[X_0^2,X_0^j,X_0^k] +[Y^2,Y^j,Y^k]
\right), \quad 
i,j,k=3,4,5.\;
\eea
This agrees exactly with \eq{gen-BH-C} if 
\be \label{f-b}
f = \cos \a,
\ee
\be \label{g0g1}
\frac{g_1}{g_0} = 1+\tan^2 \a
\ee
and 
\bea\label{NCG-M5-1}
{} [X_0^2,X_0^j,X_0^k] &=& 0,   \nn\\
{} [X_0^j, X_0^k ,X_0^l ] &=& i \Theta^{jkl}, \quad j,k,l=3,4,5, 
\eea
where
\be
\Theta^{jkl} = \e^{jkl} \frac{1}{K}\frac{C}{ (1+ C^2)^{2}}.
\ee

We note  that the relation \eq{g0g1} is precisely satisfied by the
metric \eq{G2}.  
In fact, for our $C$-field configuration, we find that 
the metric $\cG_{\m\n}$ 
is of the
form of \eq{G-M5} with
\be
g_0 = \cos^4 \frac{\a}{2},\qquad g_1 = \frac{\cos^4 \frac{\a}{2}}{\cos^2 \a}. 
\ee
It has been postulated in \cite{M2-C} that  
the metric $\cG_{\m\n}$ plays the
role of the open membrane metric just as the open string metric for
D3-brane in background $B$-field. Our analysis verifies this claim
independently.

It is  worthwhile to explore more deeply 
the meaning of the result \eq{NCG-M5-1}. 
Since the boundary variables $X^i_0$ 
satisfy precisely the mixed boundary condition of an open membrane
ending on an M5-brane with a $C$-field, one can identify $X^i_0$ as
the coordinates of the underlying M5-brane. 
The 
variables $Y^i$ should be identified with the M2-brane excitations
which describe the  protruding M5-brane as an M2-branes wedge. 
Our result \eq{NCG-M5-1} implies
that 
the M5-brane worldvolume
should satisfy the quantum
geometry relations \eq{NCG-M5-1}  for
a $C$-field configuration given by \eq{C}. The relation 
\eq{NCG-M5-1}   was obtained with a physical gauge 
$X_0 =\t, X^1= \s_1, X^9 = \s_2 +\s_{20}$. 
Properly covariantizing the results, we expect the quantum
geometry of the M5-brane takes the form
\be \label{NCG-M5}
[X_0^\m,X_0^\n,X_0^\l] = i \Theta^{\m\n\l},
\ee
where
\be
\Theta^{\m\n\l} = 
\begin{cases} 
\e^{\m\n\l} \frac{C'}{K (1- C'^2)^{2}} & \m,\n,\l =0,1,2,\\
\e^{\m\n\l} \frac{C}{K (1+ C^2)^{2}} & \m,\n,\l =3,4,5,\\
0 & \mbox{otherwise},
\end{cases}
\ee 
and $C:= -4 C_{345}$, $C':= 4 C_{012} $. 

The result \eq{NCG-M5} is intriguing. In the literature, there have been
attempts \cite{M2-C} to try to deduce the quantum geometry of the M5-brane
by following the same logic as in the D-brane case \cite{CH1} by
quantising an open membrane in the presence of a constant $C$-field.
However the analysis is much more complicated due to the nonlinear
nature of the membrane action and one can only do an approximate
analysis. Since a canonical quantization is carried out, these results
were expressed in terms of a non-vanishing commutator of 
the boundary string coordinate,
that is, in terms of a noncommutative geometry.
The expression is however quite complicated even with the simplification
due to the approximation. Compared to these results, the quantum
geometry \eq{NCG-M5} is expressed in terms of the 3-algebra and is much
simpler and more elegant. Our result suggests that the correct language
to express the quantum geometry of the M5-brane in the presence of a
constant $C$-field is in terms of a 3-bracket, rather than a commutator.

\section{Discussions}

In this paper, we have shown that the general Nahm equation which
describes a D3-brane with worldvolume $B$-field can be understood as the
boundary condition of the matrix F1-string which ends on it. This
approach provides a clear physical understanding of the 
modifications, due to the $B$-field, of the original Nahm equation: the
constant shift in the Nahm equation is due to the noncommutative
geometry of the D3-brane, and the scaling of the different components of the 
Nahm equation is due to the open string metric of the D3-brane. Applying
the same idea to the M2-M5 branes intersecting system, 
we showed that the modified Basu-Harvey equation we proposed can also be
understood in terms of the boundary condition of the multiple membranes
which end on the M5-brane if the quantum geometry of the M5-brane takes 
the form \eq{NCG-M5} and
if the open-membrane metric on the M5-brane 
is given by the metric $\cG_{\m\n}$ which appears in the nonlinear
self-duality condition of $H$. 
The prediction of the form of the quantum geometry of the M5-brane in the
presence of a constant $C$-field is the main result of this paper.

A crucial step in our proposed identification of the quantum geometry of the
M5-brane is the identification \eq{iden-H} in the presence of the $C$-field.
This is the analogue of the relation \eq{iden-F-univ} for F1-strings probing
the D1-D3 system. Now, in \eq{iden-H} we had an arbitrary scalar function of
the $C$-field, $f$, which in the system we considered in section \ref{C_BH}
was given by $f=\cos \a$, so that the relation \eq{iden-H} reads
\be \label{F-crucial}
\Ft^{ijk} = i K \cos \a [X^i,X^j, X^k]. 
\ee
An understanding of this relation can be obtained 
in the presence of the special configuration \eq{C} 
of the $C$-field. 
It is easy to see that in
this case the matrices $m$, and $\cG$ which appear in the nonlinear
self-duality condition \eq{sd-H} of $H=C+F$ are block diagonal of the
form \eq{G-M5} and hence
\be \label{HhH}
\Ht_{ijk} = f(h) H_{ijk}
\ee
for some scalar function $f(h)$. If we now substitute $H= C+F$ and
expand this relation around the given background $C$,
we obtain
\be
\Ft_{ijk} = \cos \a\; F_{ijk} + o(F^3). 
\ee
This implies that we want the identification
\be \label{Fds}
F^{ijk} = i K [X^i,X^j, X^k] + \cdots,
\ee
where $\cdots$ denotes possible higher order correction terms. 
We note that the relation \eq{Fds} is indeed what one would
expect. This can be seen by
considering a dimensional reduction of the M2-M5 system
on $X^5$. In this reduction,  it 
becomes a D2-D4 system with a 
worldvolume RR 3-form potential $C_{(3)}= C'\, dX^0 dX^1 dX^2$
and a worldvolume NS 2-form potential $B_{(2)}= - C \, dX^3 dX^4$. 
The D2-D4 system has been studied in \cite{HHM} 
and it is found that the D2-brane
tilts away from the normal of the D4-brane with an angle $\a$ due to the 
NS $B$-field. With a further T-duality on $X^1$, it 
is easy to check that our Basu-Harvey equation \eq{gen-BH-C} 
becomes the Nahm equation \eq{nahm-B-X} for the resulting 
D1-D3 system if the
3-bracket is related to the commutator through the relation:
\be \label{3-2}
[X^j, X^k, X^5] = [X^j, X^k] K^{1/2} , 
\quad j,k = 2,3,4.
\ee
Since $F^{ij5}$ is identified with $F^{ij}$ of the D3-brane theory,
the relation \eq{Fds} and \eq{3-2} leads to $F^{ij} = \l^{-2}
[X^i,X^j] +\cdots$. This relation can be mapped to one for the 
F1-strings using the $S$-duality map. The $S$-duality map
is nonlinear and in the leading order, it is
$S[F] = \Ft$. As a result we obtain precisely 
\eq{iden-F-univ} for F1-strings in the leading order. Hence we expect
\eq{F-crucial} to hold.

Our derivation suggests that both \eq{iden-F-univ} and
\eq{F-crucial} could be modified with corrections of  higher order in
$F$.  This would imply higher order corrections to 
the Basu-Harvey equation and the Nahm equation, which presumably
would describe stringy/M-theory 
corrections to the BIon description of the D1 and M2
spikes. It would be interesting to work this out in more detail.
We also comment that  for a more general configuration of $C$-field,
the nonlinear self-duality condition  will give rise to a more
complicated relation than \eq{HhH}. It would be interesting to derive the
corresponding identification \eq{F-crucial}. This 
should give a better understanding of the role of the
open membrane metric, as well as a more precise description of both the   
M5- and M2-branes in $C$-field backgrounds.

The quantum geometry expressed by \eq{NCG-M5} is intriguing. Before one can
explore its physical consequences, it is 
necessary to understand more precisely the nature of this 3-bracket and
to understand how to obtain the result \eq{NCG-M5} from a more
fundamental approach. Let us further discuss 
these issues.

On the first question, we would like to suggest that the 3-bracket is
given by a Nambu bracket \cite{nambu}. Some time ago, Nambu
advocated a new form of mechanics based on the Nambu bracket. A natural
general formulation of Nambu mechanics was analyzed in \cite{T1}. While
the usual canonical quantization is suitable for quantising the symplectic
structure of Hamiltonian mechanics, the volume preserving feature of the
Nambu bracket suggests that it is relevant for the theory of the membrane
\cite{nambu-mem}. 
We note that \eq{3-2} holds in the classical limit 
if the 3-bracket is given by a Nambu bracket
and if the commutator is given by a Poisson bracket.  
This strongly suggests that the correct form of 3-algebra to be used in 
the theory of multiple membranes is given by 
a quantization of the Nambu bracket \cite{CHMS}.

The quantization of the Nambu bracket is however a  difficult
problem. An interesting proposal using a non-associative algebra was
originally considered by Nambu \cite{nambu}. 
Deformation quantization was considered in \cite{T2}, and 
quantization in terms of cubic matrices  
was analyzed in \cite{q-nambu}, see also \cite{cubic}. 
Recently a class of Nambu brackets was
constructed \cite{CHMS} using a consistent truncation, following  
the idea of fuzzy sphere construction. An interesting property of the multiple
membrane theory based on a quantum Nambu bracket is that the entropy law
$N^{3/2}$ for multiple membranes has a natural interpretation
\cite{CHMS}, further
suggesting that the 3-algebra which is relevant for the formulation of
the theory of multiple membranes is given by a quantum Nambu bracket.

On the second question, we believe that, just like the case of D-branes,
the relation \eq{NCG-M5} can be obtained by quantising the open M2-brane
in the presence of the $C$-field. However it appears that  treating
the mixed boundary condition as a constraint and canonically quantising
the system may not be the best way to proceed. It may be possible that a
different choice of the quantisation variables and a reformulation of the
quantization is necessary. The relation of Nambu mechanics to
Hamiltonian mechanics has been explored in, for example, \cite{N-H}. It
will be interesting to explore if there is a way to reformulate the
M2-brane quantization so that the final results take
the compact form \eq{NCG-M5}. 

Now back to the possible physical consequences of \eq{NCG-M5}. In the
case of D-branes, given the noncommutative geometry expressed in terms of
a non-vanishing commutator, one immediately has a Moyal $*$-product
representation of the noncommutative geometry which allows one to
construct the noncommutative field theory as a higher derivative
non-local deformation of the original theory. This framework has led to
much interesting physics, including most notably, the IR/UV mixing
effect \cite{iruv} in noncommutative field theory, which provides a toy
model to study nonlocal effects in quantum gravity. For the present
case, it will be interesting to understand how the geometry \eq{NCG-M5}
can be realized. We think it is rather unlikely that \eq{NCG-M5} can be
realized in terms of a deformed $*$-(binary)product. On the other hand,
it appears that the content of \eq{NCG-M5} is naturally about a
deformation of a ternary operation. It has also been suggested that the 
M5-brane worldvolume theory should be non-associative \cite{ur-C}.
It will be very interesting to construct a physical model which is defined on 
a quantum space obeying  relations of the form \eq{NCG-M5} and study its
physical consequences.

It is interesting to study more details of the dimensional reduction
of the M2-M5 system to the D2-D4 system. 
In addition to \eq{3-2}, we also get from \eq{NCG-M5}
\be \label{D4-3b}
[X^0,X^1,X^2] = i \frac{C'}{K (1-C'^2)^{2}}.
\ee
This relation says that a quantum geometry
expressed in terms of a 3-bracket should appear on the D4-brane
worldvolume as a result of the presence of the RR 3-form $C_{(3)}$. 
We can also dimensionally reduce the M2-M5 system 
on the $X^1$ direction. In this case, 
the system
becomes a tilted F1-string ending on a D4-brane with worldvolume 
RR 3-form potential $C_{(3)}= -C \, dX^3 dX^4 dX^5$
and a worldvolume NS 2-form potential $B_{(2)}=  C' \, dX^0 dX^2$.
Among other things, we obtain from \eq{NCG-M5} this time the relation
\be \label{D4-3b'}
[X^3,X^4,X^5] = i \frac{C}{K (1+C^2)^{2}}.
\ee
Again a quantum geometry of the same form as \eq{D4-3b} appears due to
a RR 3-form potential.  
It is known that RR-backgrounds can give rise to nontrivial quantum
geometry in the form of nonanticommutativity \cite{nac}. More recently 
nonanticommutative geometry due to a RR 4-form potential \cite{r5} and its
AdS/CFT dual has been studied \cite{ads}. 
We emphasis our relations in
terms of a 3-bracket are different and provide a new kind of quantum
geometry due to RR-potentials.
These results are very intriguing. Since D-branes are much more under control
than M-branes, by studying the quantization of the 
D4 system in the presence of a RR 3-form, it may be possible to derive
the relations \eq{D4-3b}
and \eq{D4-3b'}
rigorously, and in turn provide us with an
understanding of the nature of the 3-bracket.

The proper understanding of the 3-algebraic structure of the quantum
geometry \eq{NCG-M5} should help us to understand and construct the
theory of the ``non-Abelian'' tensor multiplet on multiple M5-branes.
In the case of D-branes, the essential algebraic
structure, Lie-algebra, for the construction of non-Abelian gauge theory
for multiple D-branes is the same as in the theory of a single
D-brane, though in the presence of a $B$-field. We are optimistic
that something similar is true for the M5-brane(s).
 
Finally we remark that a piece of nontrivial information
about the 3-bracket can be obtained by combining our result
\eq{NCG-M5} with  an uncertainty relation proposed in \cite{ur-C}
for the  M5-brane worldvolume in the limit of large
$C$-field. The proposed relation takes the form (in our convention)
\be \label{ur}
\d X^3 \d X^4 \d X^5 \sim \frac{1}{K C}.
\ee
It is natural to
expect that the uncertainty relation \eq{ur} follows from the quantum
geometry described by \eq{NCG-M5}. If this is the case, then \eq{ur}
and \eq{NCG-M5} are in agreement if the 3-bracket obeys the scaling
\be \label{3-s}
[\cdot,\cdot,\cdot] \sim o(C^2)
\ee
in the large $C$-field limit.
This can be seen by introducing the rescaled variables
$Z^i = X^i (KC)^{1/3}$
and the  rescaled 3-bracket
$ [\cdot,\cdot,\cdot]' = \frac{1}{C^2}[\cdot,\cdot,\cdot]$,
then our relation \eq{NCG-M5} can be written as
$[Z^3,Z^4,Z^5]' =i$.
Now if $[\cdot,\cdot,\cdot]'$ is independent of $C$, and if \eq{ur}
does follow from \eq{NCG-M5}, then the
$C$-dependence in \eq{ur} is reproduced.
We note that this kind of nontrivial dependence on the
$B$-field does not occur for the $*$-commutator in the case of
noncommutative geometry of D-branes. The scaling limit \eq{3-s}
provides a constraint on the $C$-field deformation of the 3-bracket,
or, on the 3-bracket itself if the   remark in the previous
paragraph is true. 

\section*{Acknowledgements}
It is a pleasure to thank Peter Bowcock and David Fairlie for discussions and 
Pei-Ming Ho and Yutaka Matsuo for stimulating discussions and 
useful comments on the manuscript. The research of CSC is
supported by EPSRC and STFC. The research of DJS is supported by STFC.



\begin{thebibliography}{99} 

\bibitem{BLG1}
  J.~Bagger and N.~Lambert,
  ``Modeling multiple M2's,''
  Phys.\ Rev.\  D {\bf 75} (2007) 045020
  [arXiv:hep-th/0611108].

\bibitem{BLG2}
  J.~Bagger and N.~Lambert,
  ``Gauge Symmetry and Supersymmetry of Multiple M2-Branes,''
  Phys.\ Rev.\  D {\bf 77} (2008) 065008
  [arXiv:0711.0955 [hep-th]].

\bibitem{BLG3}
  J.~Bagger and N.~Lambert,
  ``Comments On Multiple M2-branes,''
  JHEP {\bf 0802} (2008) 105
  [arXiv:0712.3738 [hep-th]].

\bibitem{BLG4}
  A.~Gustavsson,
  ``Algebraic structures on parallel M2-branes,''
  arXiv:0709.1260 [hep-th].

\bibitem{BH}
  A.~Basu and J.~A.~Harvey,
  ``The M2-M5 brane system and a generalized Nahm's equation,''
  Nucl.\ Phys.\  B {\bf 713} (2005) 136
  [arXiv:hep-th/0412310].

\bibitem{LT}
  N.~Lambert and D.~Tong,
  ``Membranes on an Orbifold,''
  Phys.\ Rev.\ Lett.\  {\bf 101} (2008) 041602
  [arXiv:0804.1114 [hep-th]].
\\
J.~Distler, S.~Mukhi, C.~Papageorgakis and M.~Van Raamsdonk,
  ``M2-branes on M-folds,''
  JHEP {\bf 0805} (2008) 038
  [arXiv:0804.1256 [hep-th]].

\bm{lor}
J.~Gomis, G.~Milanesi and J.~G.~Russo,
  ``Bagger-Lambert Theory for General Lie Algebras,''
  JHEP {\bf 0806} (2008) 075
  [arXiv:0805.1012 [hep-th]].
\\
 S.~Benvenuti, D.~Rodriguez-Gomez, E.~Tonni and H.~Verlinde,
  ``N=8 superconformal gauge theories and M2 branes,''
  arXiv:0805.1087 [hep-th].
\\
P.~M.~Ho, Y.~Imamura and Y.~Matsuo,
  ``M2 to D2 revisited,''
  JHEP {\bf 0807} (2008) 003
  [arXiv:0805.1202 [hep-th]].

\bm{d2d2}
M.~A.~Bandres, A.~E.~Lipstein and J.~H.~Schwarz,
  ``Ghost-Free Superconformal Action for Multiple M2-Branes,''
  JHEP {\bf 0807} (2008) 117
  [arXiv:0806.0054 [hep-th]].
\\
J.~Gomis, D.~Rodriguez-Gomez, M.~Van Raamsdonk and H.~Verlinde,
  ``Supersymmetric Yang-Mills Theory From Lorentzian Three-Algebras,''
  JHEP {\bf 0808} (2008) 094
  [arXiv:0806.0738 [hep-th]].
\\
B.~Ezhuthachan, S.~Mukhi and C.~Papageorgakis,
  ``D2 to D2,''
  JHEP {\bf 0807} (2008) 041
  [arXiv:0806.1639 [hep-th]].

\bm{ABJM}
 O.~Aharony, O.~Bergman, D.~L.~Jafferis and J.~Maldacena,
  ``N=6 superconformal Chern-Simons-matter theories, M2-branes and their
  JHEP {\bf 0810} (2008) 091
  [arXiv:0806.1218 [hep-th]].

\bm{n6}
  J.~Bagger and N.~Lambert,
  ``Three-Algebras and N=6 Chern-Simons Gauge Theories,''
  arXiv:0807.0163 [hep-th].

\bm{m51}
P.~S.~Howe and E.~Sezgin,
  ``D = 11, p = 5,''
  Phys.\ Lett.\  B {\bf 394} (1997) 62
  [arXiv:hep-th/9611008].
\\
 P.~S.~Howe, E.~Sezgin and P.~C.~West,
  ``Covariant field equations of the M-theory five-brane,''
  Phys.\ Lett.\  B {\bf 399} (1997) 49
  [arXiv:hep-th/9702008].

\bm{m52}
 M.~Aganagic, J.~Park, C.~Popescu and J.~H.~Schwarz,
  ``World-volume action of the M-theory five-brane,''
  Nucl.\ Phys.\  B {\bf 496} (1997) 191
  [arXiv:hep-th/9701166].

\bm{m53}
P.~Pasti, D.~P.~Sorokin and M.~Tonin,
  ``Covariant action for a D = 11 five-brane with the chiral field,''
  Phys.\ Lett.\  B {\bf 398} (1997) 41
  [arXiv:hep-th/9701037].
\\
I.~A.~Bandos, K.~Lechner, A.~Nurmagambetov, P.~Pasti, D.~P.~Sorokin and M.~Tonin,
  ``Covariant action for the super-five-brane of M-theory,''
  Phys.\ Rev.\ Lett.\  {\bf 78} (1997) 4332
  [arXiv:hep-th/9701149].


\bibitem{km}
  C.~S.~Chu and E.~Sezgin,
  ``M-fivebrane from the open supermembrane,''
  JHEP {\bf 9712} (1997) 001
  [arXiv:hep-th/9710223].

\bm{CDS}
 A.~Connes, M.~R.~Douglas and A.~S.~Schwarz,
  ``Noncommutative geometry and matrix theory: Compactification on tori,''
  JHEP {\bf 9802} (1998) 003
  [arXiv:hep-th/9711162].

\bm{DH}
M.~R.~Douglas and C.~M.~Hull,
  ``D-branes and the noncommutative torus,''
  JHEP {\bf 9802} (1998) 008
  [arXiv:hep-th/9711165].


\bibitem{SW}
  N.~Seiberg and E.~Witten,
  ``String theory and noncommutative geometry,''
  JHEP {\bf 9909} (1999) 032
  [arXiv:hep-th/9908142].


\bibitem{CH1}
  C.~S.~Chu and P.~M.~Ho,
  ``Noncommutative open string and D-brane,''
  Nucl.\ Phys.\  B {\bf 550} (1999) 151
  [arXiv:hep-th/9812219].

\bibitem{CH2}
  C.~S.~Chu and P.~M.~Ho,
  ``Constrained quantization of open string in background B field and
  noncommutative D-brane,''
  Nucl.\ Phys.\  B {\bf 568} (2000) 447
  [arXiv:hep-th/9906192].

\bibitem{chu}
  C.~S.~Chu,
  ``Noncommutative open string: Neutral and charged,''
  arXiv:hep-th/0001144.

\bibitem{M2-C}
  E.~Bergshoeff, D.~S.~Berman, J.~P.~van der Schaar and P.~Sundell,
  ``A noncommutative M-theory five-brane,''
  Nucl.\ Phys.\  B {\bf 590} (2000) 173
  [arXiv:hep-th/0005026].
\\
  S.~Kawamoto and N.~Sasakura,
  ``Open membranes in a constant C-field background and noncommutative
  JHEP {\bf 0007} (2000) 014
  [arXiv:hep-th/0005123].

\bibitem{callan-mal}
  C.~G.~Callan and J.~M.~Maldacena,
  ``Brane dynamics from the Born-Infeld action,''
  Nucl.\ Phys.\  B {\bf 513} (1998) 198
  [arXiv:hep-th/9708147].


\bibitem{myers1}
  N.~R.~Constable, R.~C.~Myers and O.~Tafjord,
  ``The noncommutative bion core,''
  Phys.\ Rev.\  D {\bf 61} (2000) 106009
  [arXiv:hep-th/9911136].

\bibitem{myers2}
  N.~R.~Constable, R.~C.~Myers and O.~Tafjord,
  ``Fuzzy funnels: Non-abelian brane intersections,''
  arXiv:hep-th/0105035.

\bibitem{moriyama}
  S.~Moriyama,
  ``Noncommutative monopole from nonlinear monopole,''
  Phys.\ Lett.\  B {\bf 485} (2000) 278
  [arXiv:hep-th/0003231].

\bibitem{myers}
  R.~C.~Myers,
  ``Dielectric-branes,''
  JHEP {\bf 9912} (1999) 022
  [arXiv:hep-th/9910053].

\bibitem{KC}
  J.~L.~Karczmarek and C.~G.~.~Callan,
  ``Tilting the noncommutative bion,''
  JHEP {\bf 0205} (2002) 038
  [arXiv:hep-th/0111133].

\bibitem{HLW}
  P.~S.~Howe, N.~D.~Lambert and P.~C.~West,
  ``The self-dual string soliton,''
  Nucl.\ Phys.\  B {\bf 515} (1998) 203
  [arXiv:hep-th/9709014].

\bibitem{michishita}
  Y.~Michishita,
  ``The M2-brane soliton on the M5-brane with constant 3-form,''
  JHEP {\bf 0009} (2000) 036
  [arXiv:hep-th/0008247].

\bibitem{youm}
  D.~Youm,
  ``BPS solitons in M5-brane worldvolume theory with constant three-form
  field,''
  Phys.\ Rev.\  D {\bf 63} (2001) 045004
  [arXiv:hep-th/0009082].

\bibitem{d1}               
  D.~E.~Diaconescu,
  ``D-branes, monopoles and Nahm equations,'' 
  Nucl.\ Phys.\  B {\bf 503} (1997) 220
  [arXiv:hep-th/9608163].


\bibitem{d2}
  A.~Kapustin and S.~Sethi,
  ``The Higgs branch of impurity theories,''
  Adv.\ Theor.\ Math.\ Phys.\  {\bf 2} (1998) 571
  [arXiv:hep-th/9804027].
\\
D.~Tsimpis,
  ``Nahm equations and boundary conditions,''
  Phys.\ Lett.\  B {\bf 433} (1998) 287
  [arXiv:hep-th/9804081].

\bm{m2-BI}
 M.~R.~Garousi,
  ``On non-linear action of multiple M2-branes,''
  arXiv:0809.0985 [hep-th].
\\
  R.~Iengo and J.~G.~Russo,
  ``Non-linear theory for multiple M2 branes,''
  JHEP {\bf 0810} (2008) 030
  [arXiv:0808.2473 [hep-th]].

\bibitem{szabo}
  F.~Lizzi, N.~E.~Mavromatos and R.~J.~Szabo,
  ``Matrix sigma-models for multi D-brane dynamics,''
  Mod.\ Phys.\ Lett.\  A {\bf 13} (1998) 829
  [arXiv:hep-th/9711012].

\bibitem{BFSS}
  T.~Banks, W.~Fischler, S.~H.~Shenker and L.~Susskind,
  ``M theory as a matrix model: A conjecture,''
  Phys.\ Rev.\  D {\bf 55} (1997) 5112
  [arXiv:hep-th/9610043].

\bibitem{motl}
  L.~Motl,
  ``Proposals on nonperturbative superstring interactions,''
  arXiv:hep-th/9701025.

\bibitem{BS}
  T.~Banks, N.~Seiberg and S.~H.~Shenker,
  ``Branes from matrices,''
  Nucl.\ Phys.\  B {\bf 490} (1997) 91
  [arXiv:hep-th/9612157].

\bibitem{dvv}
  R.~Dijkgraaf, E.~P.~Verlinde and H.~L.~Verlinde,
  ``Matrix string theory,''
  Nucl.\ Phys.\  B {\bf 500} (1997) 43
  [arXiv:hep-th/9703030].

\bm{ram}
Z.~Guralnik and S.~Ramgoolam,
  ``On the polarization of unstable D0-branes into non-commutative odd
  spheres,''
  JHEP {\bf 0102} (2001) 032
  [arXiv:hep-th/0101001].

\bibitem{BC}           
  D.~S.~Berman and N.~B.~Copland,
  ``A note on the M2-M5 brane system and fuzzy spheres,'' 
  Phys.\ Lett.\  B {\bf 639} (2006) 553
  [arXiv:hep-th/0605086].
\\
C.~Krishnan and C.~Maccaferri,  
  ``Membranes on Calibrations,''
  JHEP {\bf 0807} (2008) 005 
  [arXiv:0805.3125 [hep-th]].   

\bm{metric}
J.~P.~Gauntlett and J.~B.~Gutowski,
  ``Constraining Maximally Supersymmetric Membrane Actions,''
  arXiv:0804.3078 [hep-th].
\\
 G.~Papadopoulos,
  ``M2-branes, 3-Lie Algebras and Plucker relations,''
  JHEP {\bf 0805} (2008) 054
  [arXiv:0804.2662 [hep-th]].

\bm{CHMS}
C.~S.~Chu, P.~M.~Ho, Y.~Matsuo and S.~Shiba,
  ``Truncated Nambu-Poisson Bracket and Entropy Formula for Multiple
  JHEP {\bf 0808} (2008) 076
  [arXiv:0807.0812 [hep-th]].

\bm{m2-C}
 M.~Li and T.~Wang,
  ``M2-branes Coupled to Antisymmetric Fluxes,''
  JHEP {\bf 0807} (2008) 093
  [arXiv:0805.3427 [hep-th]].

\bm{M2M5}
P.~M.~Ho and Y.~Matsuo,
  ``M5 from M2,''
  JHEP {\bf 0806} (2008) 105
  [arXiv:0804.3629 [hep-th]].
\\
 P.~M.~Ho, Y.~Imamura, Y.~Matsuo and S.~Shiba,
  ``M5-brane in three-form flux and multiple M2-branes,''
  JHEP {\bf 0808} (2008) 014
  [arXiv:0805.2898 [hep-th]].

\bm{3AlgClass}
 J.~M.~Figueroa-O'Farrill,
  ``Three lectures on 3-algebras,''
  arXiv:0812.2865 [hep-th].
\\
  P.~de Medeiros, J.~Figueroa-O'Farrill, E.~Mendez-Escobar and P.~Ritter,
  ``Metric 3-Lie algebras for unitary Bagger-Lambert theories,''
  arXiv:0902.4674 [hep-th].
\\
  P.~de Medeiros, J.~Figueroa-O'Farrill, E.~Mendez-Escobar and P.~Ritter,
  ``On the Lie-algebraic origin of metric 3-algebras,''
  arXiv:0809.1086 [hep-th].
\\
  P.~de Medeiros, J.~M.~Figueroa-O'Farrill and E.~Mendez-Escobar,
  ``Metric Lie 3-algebras in Bagger-Lambert theory,''
  JHEP {\bf 0808} (2008) 045
  [arXiv:0806.3242 [hep-th]].
\\
  P.~De Medeiros, J.~M.~Figueroa-O'Farrill and E.~Mendez-Escobar,
  ``Lorentzian Lie 3-algebras and their Bagger-Lambert moduli space,''
  JHEP {\bf 0807} (2008) 111
  [arXiv:0805.4363 [hep-th]].

\bibitem{nambu}
  Y.~Nambu,
  ``Generalized Hamiltonian dynamics,''
  Phys.\ Rev.\  D {\bf 7} (1973) 2405.

\bm{T1}
  L.~Takhtajan,
  ``On Foundation Of The Generalized Nambu Mechanics (Second Version),''
  Commun.\ Math.\ Phys.\  {\bf 160} (1994) 295
  [arXiv:hep-th/9301111].


\bm{nambu-mem}
 J.~Hoppe,
  ``On M-Algebras, the Quantisation of Nambu-Mechanics, and Volume Preserving
  Diffeomorphisms,''
  Helv.\ Phys.\ Acta {\bf 70} (1997) 302
  [arXiv:hep-th/9602020].
\\
D.~Minic,
  ``M-theory and deformation quantization,''
  arXiv:hep-th/9909022.
\\
L.~M.~Baker and D.~B.~Fairlie,
  ``Hamilton-Jacobi equations and brane associated Lagrangians,''
  Nucl.\ Phys.\  B {\bf 596} (2001) 348
  [arXiv:hep-th/0003048].
\\
Y.~Matsuo and Y.~Shibusa,
  ``Volume preserving diffeomorphism and noncommutative branes,''
  JHEP {\bf 0102} (2001) 006
  [arXiv:hep-th/0010040].
\\
B.~Pioline,
  ``Comments on the topological open membrane,''
  Phys.\ Rev.\  D {\bf 66} (2002) 025010
  [arXiv:hep-th/0201257].
\\
  T.~L.~Curtright and C.~K.~Zachos,
  ``Deformation quantization of superintegrable systems and Nambu  mechanics,''
  New J.\ Phys.\  {\bf 4} (2002) 83
  [arXiv:hep-th/0205063];
\\
T.~Curtright and C.~K.~Zachos,
  ``Classical and quantum Nambu mechanics,''
  Phys.\ Rev.\  D {\bf 68} (2003) 085001
  [arXiv:hep-th/0212267].

\bm{T2}
  G.~Dito, M.~Flato, D.~Sternheimer and L.~Takhtajan,
  ``Deformation quantization and Nambu mechanics,''
  Commun.\ Math.\ Phys.\  {\bf 183} (1997) 1
  [arXiv:hep-th/9602016].

\bibitem{q-nambu}
  H.~Awata, M.~Li, D.~Minic and T.~Yoneya,
  ``On the quantization of Nambu brackets,''
  JHEP {\bf 0102} (2001) 013
  [arXiv:hep-th/9906248].

\bm{cubic}
 Y.~Kawamura,
  Prog.\ Theor.\ Phys.\  {\bf 109} (2003) 1
  [arXiv:hep-th/0206184];
  ``Cubic matrix, Nambu mechanics and beyond,''
  Prog.\ Theor.\ Phys.\  {\bf 109} (2003) 153
  [arXiv:hep-th/0207054].


\bm{N-H}
 F.~Bayen and M.~Flato,
  ``Remarks Concerning Nambu's Generalized Mechanics,''
  Phys.\ Rev.\  D {\bf 11} (1975) 3049.
\\
N.~Mukunda and G.~Sudarshan,
  ``Relation between Nambu and Hamiltonian mechanics,''
  Phys.\ Rev.\  D {\bf 13} (1976) 2846.

\bm{iruv}
S.~Minwalla, M.~Van Raamsdonk and N.~Seiberg,
  ``Noncommutative perturbative dynamics,''
  JHEP {\bf 0002} (2000) 020
  [arXiv:hep-th/9912072].

\bm{HHM}
K.~Hashimoto and T.~Hirayama,
  ``Branes and BPS configurations of noncommutative / commutative gauge
  theories,''
  Nucl.\ Phys.\  B {\bf 587} (2000) 207
  [arXiv:hep-th/0002090].
\\
S.~Moriyama,
  ``Noncommutative monopole from nonlinear monopole,''
  Phys.\ Lett.\  B {\bf 485} (2000) 278
  [arXiv:hep-th/0003231].


\bm{ur-C}
P.~M.~Ho and Y.~Matsuo,
  ``A toy model of open membrane field theory in constant 3-form flux,''
  Gen.\ Rel.\ Grav.\  {\bf 39} (2007) 913
  [arXiv:hep-th/0701130].

\bm{nac}
H.~Ooguri and C.~Vafa,
``The C-deformation of gluino and non-planar diagrams,''
Adv.\ Theor.\ Math.\ Phys.\  {\bf 7}, 53 (2003)
[arXiv:hep-th/0302109];
%
``Gravity induced C-deformation,''
Adv.\ Theor.\ Math.\ Phys.\  {\bf 7}, 405 (2004)
[arXiv:hep-th/0303063].
\\
J.~de Boer, P.~A.~Grassi and P.~van Nieuwenhuizen,
``Non-commutative superspace from string theory,''
Phys.\ Lett.\ B {\bf 574}, 98 (2003)
[arXiv:hep-th/0302078].
\\
N.~Seiberg,
``Noncommutative superspace, N = 1/2 supersymmetry, field theory and  string
theory,''
JHEP {\bf 0306} (2003) 010
[arXiv:hep-th/0305248].
\\
N.~Berkovits and N.~Seiberg,
``Superstrings in graviphoton background and N = 1/2 + 3/2 supersymmetry,''
JHEP {\bf 0307}, 010 (2003)
[arXiv:hep-th/0306226].

\bm{r5}
S.~Ferrara, E.~Ivanov, O.~Lechtenfeld, E.~Sokatchev and B.~Zupnik,
``Non-anticommutative chiral singlet deformation of N = (1,1) gauge theory,''
[arXiv:hep-th/0405049].
\\
K.~Ito and S.~Sasaki,
  ``Non(anti)commutative N = 2 supersymmetric gauge theory from superstrings in
  the graviphoton background,''
  JHEP {\bf 0611} (2006) 004
  [arXiv:hep-th/0608143].
\\
K.~Ito, Y.~Kobayashi and S.~Sasaki,
  ``Deformation of N = 4 super Yang-Mills theory in graviphoton background,''
  JHEP {\bf 0704} (2007) 011
  [arXiv:hep-th/0612267].

\bm{ads} 
 C.~S.~Chu, S.~H.~Dai and D.~J.~Smith,
  ``AdS/CFT Duality for Non-Anticommutative Supersymmetric Gauge Theory,''
  JHEP {\bf 0805} (2008) 029
  [arXiv:0803.0895 [hep-th]].


\end{thebibliography}
\end{document}